\documentclass[a4paper,11pt]{article}
\usepackage{jcappub}
\usepackage[utf8]{inputenc}
\usepackage{amsmath, amssymb, amsthm, graphicx, epsfig, fancyhdr,epsfig, slashed, color}
\usepackage{graphicx}
\usepackage{dcolumn}
\usepackage{bm}
\usepackage{hyperref}
\usepackage[mathlines]{lineno}
\usepackage{hyperref}
\usepackage[all]{hypcap}
\hypersetup{  
    colorlinks=true,
    linkcolor=blue,
    filecolor=red,      
    urlcolor=cyan,
    citecolor=green,
    }

\newcommand{\mx}{M_{\rm X}}
\newcommand{\mbh}{M_{\rm BH}}
\newcommand{\nx}{n_{\rm X}}
\newcommand{\nxeq}{n_{\rm X}^{\rm (eq)}}
\newcommand{\Mp}{M_{P}}

\begin{document}
\title{Prospects of Indirect Detection of Dark Matter via Primordial Black Hole Induced Gravitational Waves}

\author[a]{Debarun Paul}
\emailAdd{debarun31paul@gmail.com}

\author[a,b]{Md Riajul Haque}
\emailAdd{riaj.0009@gmail.com}
\emailAdd{riaj1994@sjtu.edu.cn}

\author[a]{Supratik Pal}
\emailAdd{supratik@isical.ac.in}

\affiliation[a]{\,Physics and Applied Mathematics Unit, Indian Statistical Institute, 203 B.T. Road, Kolkata 700108, India}
\affiliation[b]{\,Tsung-Dao Lee Institute \& School of Physics and Astronomy, Shanghai Jiao Tong University,
Shanghai 201210, China}

\smallskip

\abstract{
Primordial black holes (PBHs), produced in the early Universe, can source a stochastic background of induced gravitational waves (GWs) and provide a non-thermal origin for dark matter (DM). We investigate DM production in a PBH-dominated cosmological framework, including contributions from PBH evaporation, gravitational production, and thermal freeze-in and freeze-out mechanisms, and determine the regions consistent with the observed DM relic abundance. We find that thermal freeze-in can compensate for the underabundance of PBH-sourced DM, while indirect detection remains largely insensitive due to the feeble interaction strength, making future GW observatories such as LISA and the Einstein Telescope (ET) unique probes of this scenario. For freeze-out DM, indirect detection experiments constrain regions with relatively large annihilation cross-sections, whereas GW observations probe complementary regions with heavier DM masses and smaller interaction strengths. Consequently, the same DM parameter space cannot be simultaneously probed by both indirect detection searches and GW missions. These results establish GW observations as a powerful and independent probe of DM production in PBH-dominated cosmologies, opening a new observational window into DM properties and the thermal history of the pre-BBN Universe.  
}

\maketitle


\section{Introduction}
\label{sec:intro}
Primordial black holes (PBHs) can form from high-density fluctuations in the early Universe \cite{Hawking:1974rv,Carr:1974nx,Hawking:1975vcx}, offering a powerful probe of cosmological epochs preceding big bang nucleosynthesis (BBN), where the thermal history of the Universe remains poorly constrained. Unlike astrophysical black holes, PBHs span a broad range of masses depending on the formation epoch and the underlying cosmological background, and have therefore emerged as valuable messengers of inflationary dynamics, dark matter (DM), baryogenesis, and non-standard reheating scenarios. The renewed interest in PBHs has been further strengthened by gravitational wave (GW) observations from the LIGO–Virgo–KAGRA collaboration \cite{LIGOScientific:2016aoc,LIGOScientific:2016sjg,LIGOScientific:2017bnn,LIGOScientific:2017vox,LIGOScientific:2017ycc,LIGOScientific:2017vwq}, motivating systematic studies of early Universe GW backgrounds.

The formation of PBHs is commonly associated with enhanced curvature perturbations generated during inflation. When these perturbations exceed a critical threshold at horizon reentry, gravitational collapse may occur, leading to PBH-formation with $\delta\rho/\rho \gtrsim \delta_c \sim \mathcal{O}(1)$. Several mechanisms have been proposed to generate such fluctuations, including quantum fluctuations in single-field inflation \cite{PhysRevD.48.543,PhysRevD.50.7173,Yokoyama:1998pt,Saito:2008em,Garcia-Bellido:2017mdw}, multi-field inflation \cite{Yokoyama:1995ex,Randall:1995dj,Garcia-Bellido:1996mdl,Pi:2017gih}, collapse of cosmic string loops \cite{MacGibbon:1990zk,Jenkins:2020ctp,Helfer:2018qgv,Matsuda:2005ez,Lake:2009nq}, domain wall collapse \cite{Rubin:2000dq,Rubin:2001yw}, bubble collisions during cosmological phase transitions \cite{KodamaPTP1979}, and more exotic formation channels \cite{Lu:2024xnb,Lu:2024zwa,Flores:2024lng,Flores:2023zpf,Ballesteros:2024hhq}. While the detailed formation mechanism remains model-dependent, the subsequent evolution of PBHs can be studied in a largely model-independent framework.

In the PBH-induced reheating scenario, PBHs  form during the radiation dominated era. Due to their non-relativistic nature, PBHs behave as pressure-less matter and may temporarily dominate the cosmic energy density before evaporating through Hawking radiation. The evaporation process injects entropy into the primordial plasma, reheating the Universe and initiating the radiation-dominated era \cite{Hidalgo:2011fj,Martin:2019nuw,Hooper:2019gtx,Hooper:2020evu,Hooper:2020otu,Bernal:2020bjf,Cheek:2021cfe,Cheek:2022mmy,Mazde:2022sdx,RiajulHaque:2023cqe,Calabrese:2023key,Barman:2024slw}. Within this framework, the presence of an early PBH-dominated epoch and subsequent evaporation naturally opens new avenues for particle production. In particular, DM generation through multiple mechanisms, such as Hawking evaporation, freeze-in, freeze-out, or purely gravitational processes are quite natural that may leave non-trivial effects on subsequent  phases of evolution of the Universe.  On top of that, despite overwhelming gravitational evidence for the existence of DM, its particle nature remains unknown. Extensive experimental searches including direct detection, indirect probes, and collider experiments have so far yielded null results, placing strong pressure on the conventional weakly interacting massive particle (WIMP) paradigm \cite{Arcadi:2017kky}. These constraints motivate the exploration of alternative production mechanisms and non-standard cosmological histories beyond the conventional radiation-dominated framework.

It has been shown that modifying the thermal history of the Universe, particularly during the reheating epoch, can significantly modify the existing bounds and can reopen large regions of parameter space for WIMP DM \cite{Haque:2023yra,Bernal:2022wck,Henrich:2025gsd,Henrich:2025sli}. This highlights the importance of the pre-BBN era, whose dynamics remain largely unconstrained by observations. A natural extension of this idea is DM with extremely feeble interactions with the standard model (SM). Feebly interacting massive particles (FIMPs) \cite{Bernal:2017kxu,Silva-Malpartida:2023yks} can be produced out of equilibrium through the exchange of heavy mediators \cite{Chowdhury:2018tzw,Bhattacharyya:2018evo} or via purely gravitational interactions mediated by gravitons \cite{Mambrini:2021zpp,Haque:2022kez,Clery:2022wib,Haque:2021mab}. The latter mechanism is unavoidable and provides a model-independent contribution to the DM abundance, even in the absence of any additional portal interactions. In addition, PBHs may themselves constitute a fraction of DM \cite{Belotsky:2014kca}, or act as powerful sources of DM through Hawking evaporation \cite{Gondolo:2020uqv,Sandick:2021gew,Haque:2023awl,Barman:2024slw}. More generally, DM production need not occur solely within a fully thermalized plasma. Particle production before thermalization, during epochs such as reheating, early matter domination, or PBH domination can play a crucial role, motivating a unified treatment of PBH evaporation, thermal annihilation processes, and gravitational scattering from the thermal bath.

A distinctive prediction of PBH reheating is the generation of stochastic gravitational wave backgrounds. In this scenario, two complementary sources contribute to the signal: on small scales, the discrete and Poissonian nature of the PBH distribution induces isocurvature density perturbations that act as secondary gravitational wave sources during the PBH-dominated epoch \cite{Papanikolaou:2020qtd,Domenech:2020ssp,Dalianis:2020gup,Domenech:2021ztg,Domenech:2021wkk,Papanikolaou:2021uhe,Bhaumik:2022pil,Bhaumik:2022zdd,Papanikolaou:2022chm,Bhaumik:2024qzd,Domenech:2024wao,Gross:2024wkl,Gross:2025hia}, while on larger scales PBHs behave effectively as a pressureless fluid, and the sudden transition from PBH domination to radiation domination amplifies the primordial adiabatic curvature perturbations, generating an additional gravitational wave component \cite{Inomata:2020lmk,White:2021hwi,Bhaumik:2022pil,Bhaumik:2022zdd,Bhaumik:2023wmw,Bhaumik:2024qzd,Domenech:2024wao}. As a consequence, the resulting induced gravitational wave spectrum exhibits characteristic doubly peaked features whose amplitudes and peak frequencies encode fundamental information about the PBH population, including the PBH mass, the initial energy fraction at formation, the background equation of state etc. In a previous work by the same authors \cite{Paul:2025kdd}, it was demonstrated that these doubly peaked induced gravitational wave signals associated with PBH reheating provide a powerful probe of the very early Universe. In particular, the spectral structure allows one to reconstruct key PBH parameters such as the formation mass and the initial abundance  through signal-to-noise ratio analyses. The corresponding parameter uncertainties were estimated using the Fisher matrix formalism, incorporating realistic instrumental noise curves of future gravitational wave observatories, including the Laser Interferometer Space Antenna (LISA)~\cite{amaroseoane2017laser,Baker:2019nia} and the Einstein Telescope (ET)~\cite{Punturo_2010,Hild:2010id}. These studies revealed that next-generation interferometers possess significant sensitivity to PBH-induced reheating and offer a unique opportunity to probe the expansion history of the Universe well before the onset of BBN. 

Motivated by these findings, an important question naturally arises: whether PBH reheating can also leave observable imprints in the DM sector. In particular, it is of interest to investigate whether DM produced during the PBH-dominated epoch may give rise to detectable signals in indirect searches, and to assess the corresponding detection prospects in upcoming GW experiments as well as in DM indirect detection searches. Importantly, these two observational avenues probe complementary aspects of the PBH reheating era: while induced gravitational waves encode information about the early Universe expansion history and PBH dynamics, indirect DM searches are sensitive to the annihilation properties of DM particles. Accordingly, we present a comprehensive investigation of the phenomenological interplay between DM production, induced gravitational waves, and indirect detection probes within the PBH reheating framework. We analyse the combined impact of PBH evaporation, thermal annihilation processes, and gravitational scattering mediated by graviton exchange on the resulting DM relic abundance. The parameter space consistent with the observed DM density is subsequently confronted with current indirect detection limits derived from $\gamma$-ray \cite{MAGIC:2016xys,MAGIC:2021mog,HESS:2021zzm,HESS:2015cda,HAWC:2018eaa,HAWC:2017mfa,Thorpe-Morgan:2020czg,Hoof:2018hyn,CherenkovTelescopeArray:2023aqu} and neutrino observations \cite{IceCube:2023ies,IceCube:2017rdn,Super-Kamiokande:2020sgt,Frankiewicz:2015zma}. By mapping these constraints onto the predicted gravitational wave spectra, we identify the regions of parameter space accessible to future interferometers such as LISA and the ET.

Our results show that, although both indirect detection experiments and GW observations probe DM properties in PBH-dominated cosmologies, they are sensitive to distinct and non-overlapping regions of the DM parameter space in the freeze-out scenario. In particular, indirect detection searches primarily constrain regions with relatively large annihilation cross-sections and lower DM masses, whereas GW observations probe heavier DM masses and smaller annihilation cross-sections through their impact on the early Universe expansion history. As a consequence, the same DM parameter space in the $(\mx,\langle\sigma v\rangle)$ plane cannot be simultaneously probed by both indirect detection observations and GW missions such as LISA and ET. Nevertheless, GW observations provide a unique and powerful probe of otherwise inaccessible regions of parameter space, offering an independent observational window into DM production and the thermal history of the pre-BBN Universe.


\section{Formation of PBH in the early Universe}
\label{sec:PBH_domination}

\subsection{Basic formalism of PBH reheating}\label{subsec:PBH_general}

PBHs can form in the early Universe from the gravitational collapse of enhanced density fluctuations. In this work, we focus on PBH-formation occurring during a radiation-dominated (RD) epoch. In such a background, overdense regions that re-enter the Hubble horizon may collapse into black holes (BHs) provided the density contrast exceeds a critical threshold. The mass of the resulting PBH is expected to be of the order of the total energy enclosed within the Hubble volume at the time of formation. Under these assumptions, the PBH-formation mass $M_{\rm in}$ can be written as
\begin{equation}
M_{\rm in}
=
\gamma \rho_{\rm R}(t_{\rm in})
\frac{4}{3}\pi
\frac{1}{H_{\rm in}^{3}}
=
4\pi\gamma
\frac{\Mp^2}{H_{\rm in}}
\simeq
1.3~\gamma~{\rm g}
\left(
\frac{10^{14}~{\rm GeV}}{H_{\rm in}}
\right),
\label{Eq:min}
\end{equation}
where $H_{\rm in}$ and $\rho_{R}(t_{\rm in})$ denotes the Hubble parameter, and radiation energy density at the time of PBH-formation, respectively.  $\gamma$ is the collapse efficiency parameter, which quantifies the fraction of the horizon mass that collapses into the BH. More refined analyses of PBH-formation incorporate the effects of the background cosmology as well as the detailed profile of scalar perturbations~\cite{Musco:2012au,Musco:2008hv,Hawke:2002rf,Niemeyer:1997mt,Escriva:2021pmf,Escriva:2019nsa,Escriva:2020tak,Escriva:2021aeh}. However, a complete analytical understanding of the collapse efficiency parameter is still lacking. In the present analysis, we adopt the standard value of the collapse efficiency parameter, $\gamma=(\frac13)^\frac32\simeq 0.2$ \cite{Carr:1974nx}. The Hubble rate at formation is related to the radiation energy density according to

\begin{equation}
H_{\rm in}
=
\sqrt{\frac{\rho_R(t_{\rm in})}{3\Mp^2}}
=
\sqrt{\frac{\alpha}{3}}\,
\frac{T_{\rm in}^2}{\Mp},
\label{Eq:hin}
\end{equation}
where $\alpha \equiv g_\ast(T)\pi^2/30$, and $g_\ast(T)$ denotes the effective number of relativistic degrees of freedom at temperature $T$. The quantity $T_{\rm in}$ corresponds to the radiation temperature at the time of PBH-formation.
\begin{figure}
    \centering
    \includegraphics[scale=0.35]{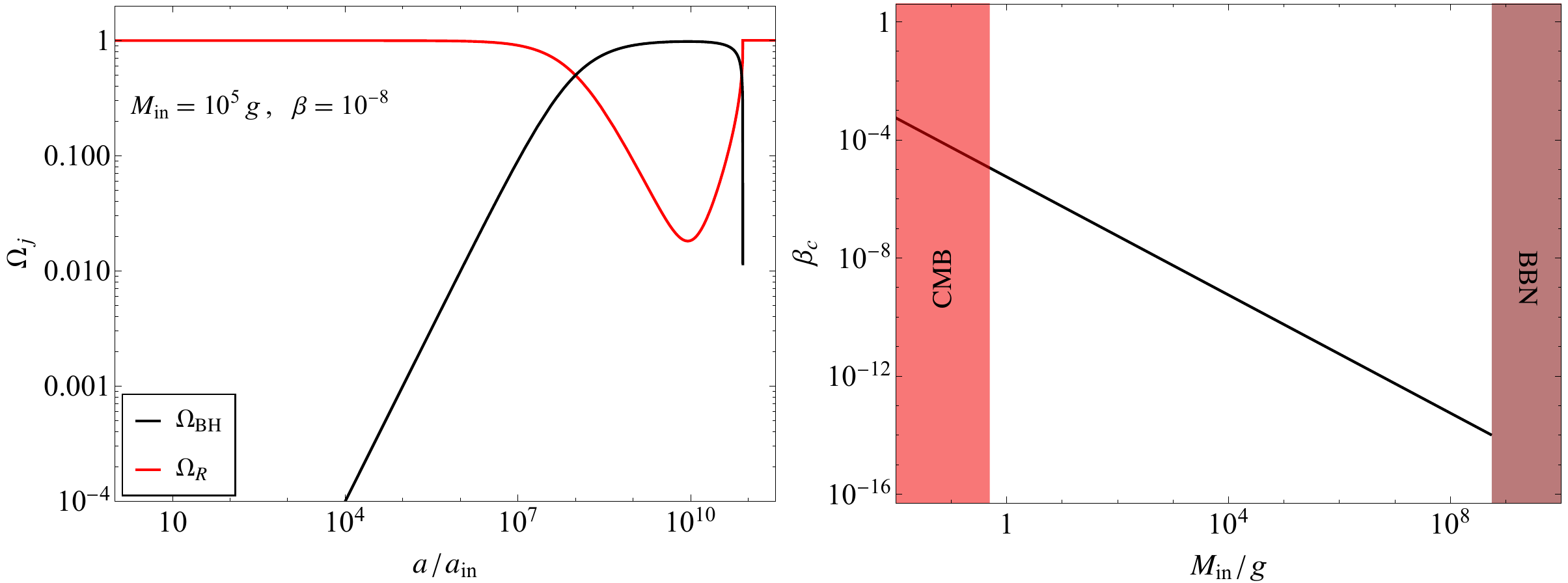}
    \caption{\it \textbf{Left panel:} Evolution of the normalized energy densities,$\Omega_j \equiv \rho_j /(3 \Mp^2 H^2)$, as functions of the normalized scale factor. \textbf{Right panel:} Dependence of the critical PBH energy fraction $\beta_c$ on the initial PBH mass $M_{\rm in}$. The shaded regions indicate excluded parameter space: the red band corresponds to constraints from CMB on the PBH-formation mass, while the brown band represents bounds arising from BBN.}
    \label{fig:betac}
\end{figure}

Since PBHs form after inflation, the Hubble scale at formation must satisfy $H_{\rm in}\lesssim H_{\rm e}$, where $H_{\rm e}$ is the Hubble parameter at the end of inflation. Current cosmic microwave background (CMB) observations constrain this scale to $H_{\rm e}\lesssim 10^{14}~{\rm GeV}$~\cite{Planck:2018jri}. As a result, Eq.~\eqref{Eq:min} implies a lower bound on the PBH-formation mass, $M_{\rm in}\gtrsim \mathcal{O}(1~{\rm g})$, for $\gamma\sim\mathcal{O}(1)$. Combining Eqs.~\eqref{Eq:min} and~\eqref{Eq:hin}, the radiation temperature at the time of PBH-formation is given by
\begin{equation}
T_{\rm in}
=
\left(
\frac{4\pi\gamma}{\sqrt{\alpha/3}}
\right)^{1/2}
\frac{\Mp^{3/2}}{M_{\rm in}^{1/2}}
\simeq
9.7\times10^{15}
\sqrt{\gamma}
\left(
\frac{1~{\rm g}}{M_{\rm in}}
\right)^{1/2}
~{\rm GeV}.
\label{Eq:tin}
\end{equation}

The subsequent cosmological evolution of PBHs is governed by three parameters: the formation mass $M_{\rm in}$, the initial PBH energy fraction $\beta$, and the background equation-of-state parameter $w$ at the time of formation. In the present work, we restrict our attention to PBHs forming during a radiation-dominated epoch and therefore set $w=\frac{1}{3}$, throughout our analysis. The cosmological evolution of the different energy components is governed by the coupled Boltzmann equations. Expressed in terms of the scale factor, these equations take the form

\begin{equation}
\frac{\mathrm{d} \rho_{\rm R}}{\mathrm{d} a}
+
4\frac{\rho_{\rm R}}{a}
=
-
\frac{\rho_{\rm BH}}{M_{\rm BH}}
\frac{\mathrm{d} M_{\rm BH}}{\mathrm{d} a},
\end{equation}
\begin{equation}
\frac{\mathrm{d} \rho_{\rm BH}}{\mathrm{d} a}
+
3\frac{\rho_{\rm BH}}{a}
=
\frac{\rho_{\rm BH}}{M_{\rm BH}}
\frac{\mathrm{d} M_{\rm BH}}{\mathrm{d} a},
\end{equation}
\begin{equation}
\frac{\mathrm{d} M_{\rm BH}}{\mathrm{d} a}
=
-
\epsilon
\frac{\Mp^4}{M_{\rm BH}^2}
\frac{1}{aH}.
\label{eq:pbh_mass}
\end{equation}
Here $\rho_{\rm BH}$ denote the energy density of the PBHs.  The parameter $\epsilon \equiv 3.8 \left( \frac{\pi}{480} \right) g_\ast(T_{\rm BH})$ encodes the total number of particle species emitted during Hawking evaporation, where \(g_\ast(T_{\rm BH})\) represents the effective number of relativistic degrees of freedom at the BH temperature. The numerical factor \(3.8\) accounts for the greybody corrections to the Hawking spectrum~\cite{Arbey:2019mbc,Cheek:2021odj,Baldes:2020nuv}. The evolution of the PBH energy density is governed by two competing effects: dilution due to the expansion of the Universe and mass loss through Hawking radiation. The latter is described by Eq.~\eqref{eq:pbh_mass}, whose solution yields the PBH mass at any cosmic time \footnote{Note that, thermal absorption in the early Universe can significantly modify the PBH mass, lifetime, reheating, and dark matter production, we neglect this effect throughout this analysis \cite{Haque:2026vvp,Kallifatides:2026sik}.
},
\begin{equation}
M_{\rm BH}(t)
=
M_{\rm in}
\left[
1
-
\Gamma_{\rm BH}
(t - t_{\rm in})
\right]^{1/3},
\end{equation}
where the total decay width associated with the PBH can be written as 
\begin{equation}
\Gamma_{\rm BH}
=
\frac{3\epsilon \Mp^4}{M_{\rm in}^3}.
\end{equation}
The total lifetime of a PBH is therefore given by $
t_{\rm ev} = \Gamma_{\rm BH}^{-1}$. Figure~\ref{fig:betac} shows the evolution of the different energy components as a function of the normalized with the scale factor at the point of formation, $a/a_{\rm in}$. The figure  indicates the existence of a threshold value of the initial PBH energy fraction, denoted by $\beta_{\rm c}$. When the initial abundance satisfies $\beta > \beta_{\rm c}$, the energy density of PBHs  overtakes that of the radiation prior to the completion of PBH evaporation. In this regime, the Universe undergoes an intermediate PBH-dominated phase, and the subsequent reheating of the Universe is driven predominantly by Hawking radiation from evaporating PBHs. We refer to this cosmological history as \emph{PBH reheating}.
The critical value is given by~\cite{RiajulHaque:2023cqe, Haque:2023awl, Haque:2024eyh}
\begin{equation}
\beta_{\rm c} = \sqrt{\frac{3\, \epsilon}{8\pi\, \gamma}} \left( \frac{\Mp}{M_{\rm in}} \right)
\sim 7.3 \times 10^{-6} \left( \frac{1\, \mathrm{g}}{M_{\rm in}} \right) \left( \frac{0.2}{\gamma} \right)^{1/2},
\end{equation}
which determines whether PBHs can temporarily dominate the cosmic energy budget before their complete evaporation. The completion of PBH evaporation corresponds to a reheating of the Universe. The radiation temperature at the end of evaporation is given by
\begin{equation}
T_{\rm ev}
=
\left(
\frac{360\,\epsilon^2}
{\pi^2\,g_\ast(T_{\rm ev})}
\right)^{1/4}
\left(
\frac{\Mp}{M_{\rm in}}
\right)^{3/2}
\Mp
\simeq
2.7\times10^{10}
\left(
\frac{1~{\rm g}}{M_{\rm in}}
\right)^{3/2},
\label{eq:Teva}
\end{equation}
where we have adopted $g_\ast(T_{\rm ev})=106.75$ and $g_\ast(T_{\rm BH})=108$. Consistency with the BBN  requires that the Universe be radiation dominated by the onset of BBN, which implies  $T_{\rm ev} \geq T_{\rm BBN} \simeq 4~{\rm MeV}$ \cite{Kawasaki:1999na,Kawasaki:2000en,Hasegawa:2019jsa}.
This condition places an upper bound on the PBH-formation mass. Combining CMB and BBN constraints, the allowed mass range is therefore
\begin{equation}
0.5~{\rm g}
\left(\frac{\gamma}{0.2}\right)
\lesssim
M_{\rm in}
\lesssim
4.8\times10^{8}~{\rm g}
\left(\frac{g_\ast(T_{\rm ev})}{106.75}\right)^{-1/6}
\left(\frac{g_\ast(T_{\rm BH})}{108}\right),
\end{equation}
which we adopt throughout this work. In the next section, we investigate the relic abundances of particles produced through PBH evaporation. We then connect these relics to DM phenomenology and determine the allowed parameter space for PBH-induced DM production within the PBH reheating scenario.

\subsection{Particle production  from PBH evaporation: Calculation of DM relics}
\label{subsec:DM_production_PBH}

The production rate of any particle from the evaporation of the PBH depends on the mass and spin of the PBH. For simplicity, we restrict our analysis to non-rotating (Schwarzschild) PBHs. For a given particle species, \textit{j}, if the total number of emitted particle per PBH is $\mathcal{N}_j$, the Hawking emission spectrum at energy $E$, can be expressed as~\cite{Cheek:2021odj} 
\begin{eqnarray}
    \frac{{\rm d}^2{\mathcal N_j}}{{\rm d}{E}{\rm d}{t}} = \frac{27g_j G^2\mbh^2}{2\pi}\frac{\psi_{s_j}(E)(E^2-\mu_j^2)}{\exp(E/T_{\rm BH})-(-1)^{2s_j}}\,,
\end{eqnarray}
where $g_j$, $s_j$, and $m_j$ denote the internal degrees of freedom, spin, and mass of the species $j$, respectively. The factor $\psi_{s_j}(E)$ is the absorption cross section which is normalized to the geometric optics limit. $G$ is the Newtonian gravitational constant, related to the reduced Planck mass as $\Mp\equiv 1/\sqrt{8\pi G}$. $T_{\rm BH}$ presents temperature of BH, which is related to mass of the BH, $\mbh$, as $T_{\rm BH}\equiv \frac{\Mp^2}{\mbh}$. Introducing two dimensionless parameters $x\equiv E/T_{\rm BH}$ and $z_j\equiv m_j/T_{\rm BH}$, the energy-integrated emission rate of the BH can be defined as
\begin{eqnarray}
\label{eq:Hwtotrate}
\Gamma_{{\rm BH} \to j}
\equiv
\int {\rm d}p\,
\frac{{\rm d}^2 \mathcal{N}_{j}}{{\rm d}p\,{\rm d}t},
\end{eqnarray}
which represents the total emission rate of particle species $j$ from a BH. 
For Schwarzschild BHs, this quantity admits an analytic expression~\cite{Cheek:2021odj}
\begin{eqnarray}
\label{eq:Gamma_DM}
\Gamma_{{\rm BH} \to j}
&=&
\frac{27 g_s}{512\pi^4}
\frac{\varepsilon_j}{G M_{\rm BH}}
\left[
z_j\,{\rm Li}_2(\varepsilon_j e^{-z_j})
+
{\rm Li}_3(\varepsilon_j e^{-z_j})
\right],
\end{eqnarray}
where ${\rm Li}_n$ denotes the polylogarithm function of order $n$, and
$\varepsilon_j \equiv (-1)^{2 s_j}$, with $s_j$ being the spin of the emitted particle. To estimate the evolution of DM number density produced via PBH evaporation, we employ the Boltzmann equation,
\begin{eqnarray}
\label{eq:ndm_evol}
\frac{{\rm d} n_{\rm X}}{{\rm d} a}
+
3\frac{n_{\rm X}}{a}
=
\Gamma_{{\rm BH} \to X}
\frac{\rho_{\rm BH}}{M_{\rm BH}}
\frac{1}{a H},
\end{eqnarray}
where the Hubble expansion rate is given by $H^2
=
\frac{1}{3 \Mp^2}
\left(
\rho_{\rm R}
+
\rho_{\rm BH}
+
\mx n_{\rm X}
\right)
\simeq
\frac{1}{3 \Mp^2}
\left(
\rho_{\rm R}
+
\rho_{\rm BH}
\right)$, 
with the last equality following from the subdominant contribution of DM energy density. In the context of the PBH reheating scenario, the present-day relic abundance of  DM  species $X$, evaluated at the temperature $T_0$, can be expressed as~\cite{Mambrini:2021cwd} 
\begin{eqnarray}
\Omega_X h^2
=
1.6 \times 10^8\,
\frac{g_0}{g_{\rm RH}}\,
\frac{  n_{\rm X}(a_{\rm ev})}{T_{\rm ev}^3}\,
\frac{\mx}{\rm GeV}\,,
\label{Eq:omegah2}
\end{eqnarray}
with $a_{\rm ev}$ being the scale factor at the time of PBH evaporation. The effective numbers of relativistic degrees of freedom for entropy at the end of reheating and at present are
$g_{\rm RH} = 106.75$ and $g_0 = 3.91$, respectively.
We assume that the effective degrees of freedom for entropy and radiation are identical. Note that, since reheating is achieved through PBH evaporation with PBH-domination, the scale factor at evaporation coincides with that at the end of reheating. Therefore, we set $a_{\rm ev} = a_{\rm RH}$, where $a_{\rm RH}$ is the scale factor at the end of reheating. Instead of performing a full numerical evolution, one can estimate the DM  number density by computing the total number of particles emitted over the lifetime of a PBH  denoted by $N_X$. The resulting DM number density is then given by
\begin{equation}
n_{\rm X} = N_X \times n_{\rm BH},
\end{equation}
where $n_{\rm BH} = \rho_{\rm BH}/M_{\rm BH}$ denotes the PBH number density. To do that it is convenient to distinguish two qualitatively different regimes depending on the relative magnitude of the particle mass and the BH temperature at formation, $T_{\rm BH}^{\rm in}$~\footnote{For the Schwarzchild BHs, the BH temperature at the time of its formation, $T_{\rm BH}^{\rm in}$, is related to $M_{\rm in}$ as $T_{\rm BH}^{\rm in}\simeq 10^{13}\left(\frac{1\,{\rm g}}{M_{\rm in}}\right)$~GeV.}. If the emitted particle is light compared to the initial BH  temperature, $m_j \lesssim T_{\rm BH}^{\rm in}$, emission remains kinematically allowed throughout PBH-lifetime. Integrating the emission rate in Eq.~\eqref{eq:Hwtotrate} from the formation time $t_{\rm in}$ to the evaporation time $t_{\rm ev}=\Gamma_{\rm BH}^{-1}$, one obtains the total number of particles emitted per PBH as
\begin{equation}
\mathcal{N}_j^{m_j \lesssim T_{\rm BH}^{\rm in}}
= \int_{t_{\rm in}}^{t_{\rm ev}} {\rm d}t \,
\frac{{\rm d}\mathcal{N}_j}{{\rm d}t}
\simeq \frac{15\, g_j\, \zeta(3)}{g_\ast \pi^4}\,
\frac{M_{\rm in}^2}{\Mp^2}
\;\simeq\;10^8\left(\frac{M_{\rm in}}{1\,{\rm g}}\right)^2 ,
\label{eq:Mj_light}
\end{equation}
where $\zeta(3)$ denotes the Riemann zeta function. This expression shows that, for sufficiently light species, the total yield scales quadratically with the initial PBH mass and is independent of the particle mass. 

Conversely, if the particle mass exceeds the initial BH temperature, $m_j \gtrsim T_{\rm BH}^{\rm in}$, particle emission becomes efficient only at later times, when the PBH temperature increases to $T_{\rm BH}\simeq m_j$. In this case, the time integration begins at $t_j$, defined through $T_{\rm BH}(t_j)=m_j$, which yields
\begin{equation}
t_j = \Gamma_{\rm BH}^{-1}
\left(1-\frac{\Mp^6}{m_j^3 M_{\rm in}^3}\right).
\label{eq:tj_def}
\end{equation}
Integrating the emission rate from $t_j$ to $t_{\rm ev}$, the total number of emitted particles is then given by
\begin{equation}
\mathcal{N}_j^{m_j \gtrsim T_{\rm BH}^{\rm in}}
= \int_{t_j}^{t_{\rm ev}} {\rm d}t \,
\frac{{\rm d}\mathcal{N}_j}{{\rm d}t}
\simeq \frac{15\, g_j\, \zeta(3)}{g_\ast \pi^4}\,
\frac{\Mp^2}{m_j^2}
\;\simeq\;10^{14}\left(\frac{10^{10}\,{\rm GeV}}{m_j}\right)^2 ,
\label{eq:Mj_heavy}
\end{equation}
with an additional factor of $3/4$ for fermionic species.
From Eqs.~\eqref{eq:Mj_light} and \eqref{eq:Mj_heavy}, it is evident that the scaling of the total particle yield differs qualitatively in the two regimes: for $m_j \lesssim T_{\rm BH}^{\rm in}$, the yield depends solely on the initial PBH mass, $\mathcal{N}_j \propto M_{\rm in}^2$, whereas for $m_j \gtrsim T_{\rm BH}^{\rm in}$ it becomes independent of $M_{\rm in}$ and instead scales as $\mathcal{N}_j \propto m_j^{-2}$.

\begin{figure}
    \centering
    \includegraphics[scale=0.35]{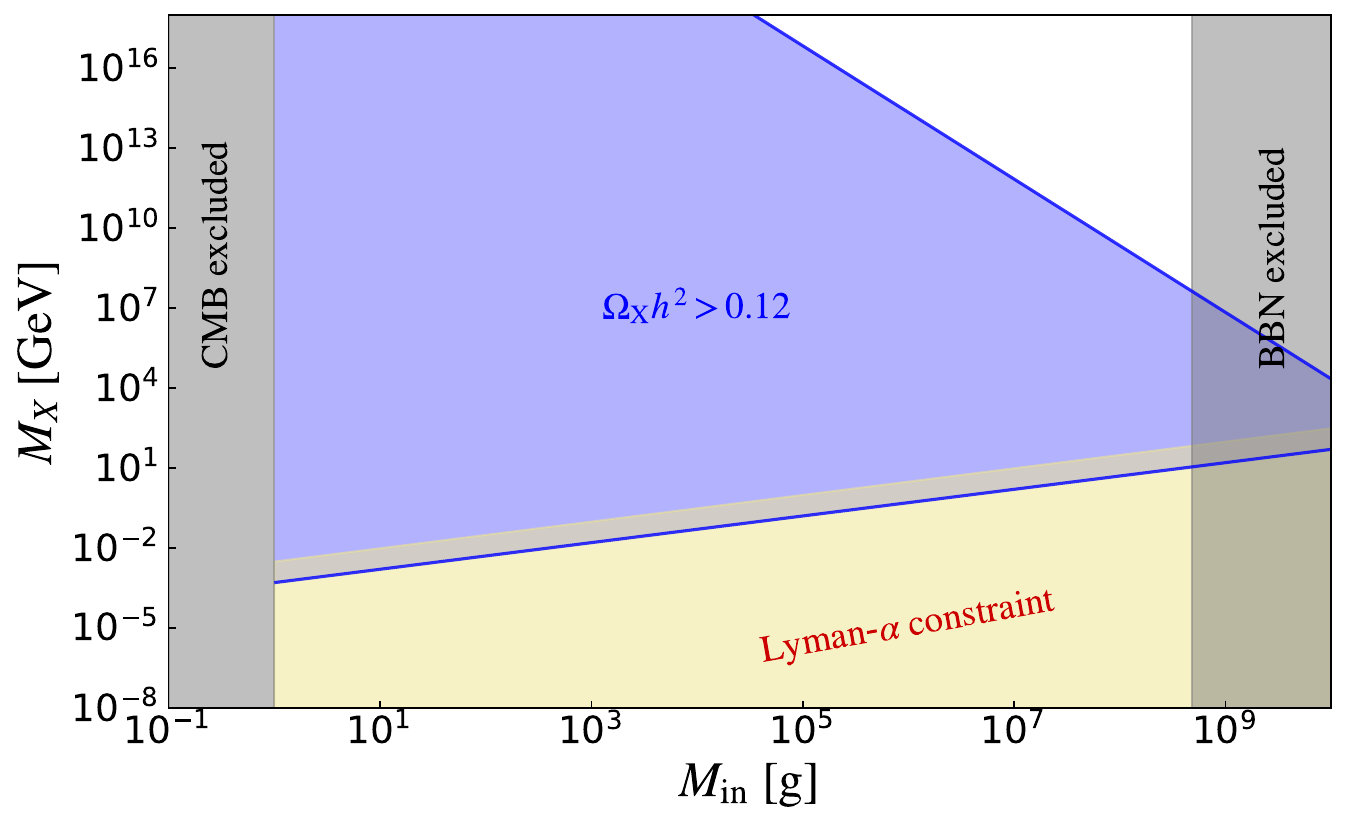}
    \caption{\it Allowed parameter range in the $M_{\rm in}$-$\mx$ plane, considering the production of DM only from the decay of PBH. The blue shaded region is excluded due to the overproduction of DM. The grey-hatched areas are ruled out by BBN and CMB constraints. The excluded region from Lyman-$\alpha$ observation is shown by light orange region. Figure reproduced from Ref.~\cite{Haque:2023awl}.}
    \label{fig:MBH_MX_onlyPBH}
\end{figure}

In addition to abundance considerations, DM produced from PBH evaporation must also satisfy structure formation constraints. In particular, observations of the Lyman-$\alpha$ forest constrain the free-streaming properties of DM, thereby imposing a lower bound on its mass. These bounds arise from the requirement that DM should not be excessively warm, as excessive free streaming would suppress small-scale structures inconsistent with observations. Recent analyses of the Lyman-$\alpha$ forest data, combining HIRES/MIKE and XQ-100 measurements, place a lower limit on the mass of warm DM (WDM) $m_{\rm WDM}>5.3$ keV at $2\sigma$ C.L.~\cite{Irsic:2017ixq}.
The impact of these bounds is governed by the velocity of DM, which is strongly correlated with its mass. For a thermal WDM particle, the present-day velocity can be parametrized as~\cite{Bode:2000gq}
\begin{equation}
v_{\rm WDM} \simeq 3.9\times10^{-8}
\left(\frac{\mathrm{keV}}{m_{\rm WDM}}\right)^{4/3},
\end{equation}
indicating that lighter DM particles possess larger velocity dispersion and hence stronger suppression of small-scale structure.

During a PBH-dominated epoch ($\beta>\beta_c$), DM particles are produced non-thermally with momentum set by Hawking evaporation. The present-day velocity of such DM particles is given by $v_0 = p_0/M_{\rm X}$, where the redshifted momentum, obtained using entropy conservation, is
\begin{equation}
p_0 =
\left(\frac{g_{s,{\rm eq}}}{g_{s,{\rm ev}}}\right)^{1/3}
\frac{T_{\rm eq}}{T_{\rm ev}}
\frac{\Omega_R^{(0)}}{\Omega_m^{(0)}}\, p_{\rm ev},
\end{equation}
with $\Omega_R^{(0)}\simeq 5.4\times 10^{-5}$ and $\Omega_m^{(0)}\simeq 0.315\pm 0.007$, being the present-day abundance of radiation and matter, respectively, according to Planck-18 data~\cite{Planck:2018vyg}. $g_{s,{\rm ev}}$ and $g_{s,{\rm eq}}$ are the effective degree of freedom for entropy at the time of matter-radiation equality and PBH-evaporation, respectively. $T_{\rm eq}$ presents the temperature at the time of matter-radiation equality, which is $0.8$ eV~\cite{Planck:2018vyg}. For the light DM \textit{i.e.} $\mx\ll T_{\rm BH}^{\rm in}$, its momentum at the time of its production ($p_{\rm ev}$) is driven by the formation temperature of the PBH \textit{i.e.} $p_{\rm ev}\sim T_{\rm BH}^{\rm in}$.
Since PBH-produced DM can inherit significantly larger momenta than thermal WDM, their free-streaming length is correspondingly enhanced, leading to a stronger lower bound on the DM mass.

As a result, the WDM constraint can be translated into an approximate lower bound on the mass of DM for PBH-dominated reheating ($\beta>\beta_c$) of the form
\begin{equation}
M_{\rm X}\;\gtrsim\;
7\times10^{-7}
\left(\frac{m_{\rm WDM}}{\mathrm{keV}}\right)^{4/3}
\left(\frac{M_{\rm in}}{M_{P}}\right)^{1/2}
\mathrm{GeV},
\end{equation}
where $M_{\rm in}$ denotes the initial PBH mass. This bound implies that for lighter PBHs, the minimum allowed DM mass can lie well above the conventional keV scale, excluding a portion of the otherwise viable parameter space inferred from relic abundance alone.

Figure~\ref{fig:MBH_MX_onlyPBH} illustrates the allowed region in the $M_{\rm in}$-$M_X$ plane for DM production purely from PBH evaporation. The grey-hatched regions denote excluded regions by BBN and CMB. The blue-shaded region present the overproduction of DM from PBH. The light orange region denotes the excluded region from the Lyman-$\alpha$ observation. For $\mx\lesssim T_{\rm BH}^{\rm in}$, the abundance of DM decreases with increasing $M_{\rm in}$ due to the inverse scaling of the emission rate (cf. Eq.~\eqref{eq:Gamma_DM}). In contrast, for $\mx\gtrsim T_{\rm BH}^{\rm in}$, the yield decreases with increasing DM mass, reflecting the $\mx^{-2}$ scaling of the emitted particle number (cf. Eq.~\eqref{eq:Mj_heavy}). The remaining region corresponds to an underabundant DM population, motivating the consideration of additional production mechanisms discussed in the following sections.

\section{Additional production mechanism of DM}
\label{sec:DM_GW}
Apart from production through PBH evaporation, DM can also be generated through interactions with the thermal plasma in the early Universe. In particular, DM can be produced purely via gravitational interactions with the thermal bath, a mechanism commonly referred to as \emph{gravitational freeze-in}~\cite{Nanopoulos:1983up,Ellis:1983ew}. This contribution is unavoidable, as gravity universally couples to the energy-momentum tensor of all particle species. In addition, if DM possesses non-gravitational interactions with the thermal bath, it may also be produced via standard freeze-in or freeze-out mechanisms. In the following, we discuss these production channels and assess their impact on the resulting DM abundance.

\subsection{Gravitational production of DM}
\label{sec:grav_DM}

Even in the absence of any non-gravitational interactions, DM is inevitably produced through gravitational scatterings involving particles in the thermal bath. Gravitational production proceeds dominantly through $2\to2$ scattering process, mediated by graviton exchange in the $s$-channel. Owing to the Planck-suppressed nature of the interaction, DM never attains thermal equilibrium with the plasma, and its abundance is generated gradually as the Universe cools. As a result, the production is most efficient at the highest temperatures attained after inflation and is therefore sensitive to the reheating temperature $T_{\rm RH}$.

\begin{figure}
    \centering
    \includegraphics[scale=1.0]{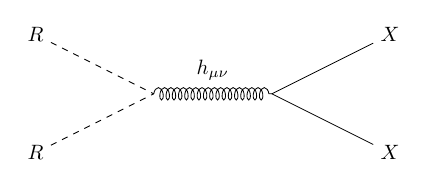}
    \caption{\it s-channel Feynman diagram for the production of DM from scattering of the SM particles through the exchange of graviton. $R$ and $X$ are the radiation and DM particles, respectively. $h_{\mu\nu}$ presents the graviton field.}
    \label{fig:feynman}
\end{figure}

The production rate per unit volume can be parametrized as $R(T)=\delta \, \frac{T^8}{M_P^4},$~\cite{Bernal:2018qlk} where $\delta$ is a numerical coefficient encoding the spin and degrees of freedom of DM.
For scalar DM, one finds $\delta \simeq 1.9 \times 10^{-4}$, while for fermions, it is $1.1\times10^{-3}$~\cite{Bernal:2018qlk}. The evolution of the DM number density is governed by
\begin{equation}
\label{eq:grav_BE}
\frac{d(n_X a^3)}{da} = \frac{R(T)}{a^2 H}.
\end{equation}
The evolution of $\nx$ is strongly affected by the post-inflationary entropy injection and thereby, on the post-inflationary PBH-domination. Thus, we elaborate both the scenarios: \textit{i.e.} without PBH-domination ($\beta<\beta_c$), and the case for PBH-domination ($\beta>\beta_c$).
\paragraph{Without PBH domination ($\beta<\beta_c$):} 
Assuming that reheating occurs sufficiently early through the decay of the inflaton, with a reheating temperature satisfying $T_{\rm RH} \gg \mx$, the Universe enters a radiation-dominated (RD) phase after inflaton decay. During RD era, the Hubble parameter can be parametrized as $H(a) = H_{\rm RH} \left(a_{\rm RH}/a \right)^2$, where $H_{\rm RH}$
denotes the Hubble rate at the end of reheating. Within this framework, we integrate Eq.~\eqref{eq:grav_BE} from $a_{\rm RH}$ to a later scale factor $a$ during radiation domination. The resulting DM number density reads
\begin{equation}
n_{\rm X}(a)=\delta \sqrt{\frac{10}{g_*}}\,
\frac{T_{\rm RH}^6\,a_{\rm RH}^3}{\pi M_{P}^3\,a^3}
\left[1-\left(\frac{a_{\rm RH}}{a}\right)^3\right]
\simeq
\delta \sqrt{\frac{10}{g_*}}\,
\frac{T_{\rm RH}^3\,T^3}{\pi M_{P}^3},
\end{equation}
After reheating, we assume that no significant entropy injection occurs, implying that the yield of DM, defined as the ratio of the number density to the entropy density, remains conserved until the present epoch. This assumption no longer holds in scenarios with PBH domination, which will be discussed below. For $\beta < \beta_{\rm c}$, corresponding to the absence of PBH domination, the present-day yield is therefore given by
\begin{equation}
Y_0 \equiv \frac{n_{\rm X}(T)}{s(T)} \simeq
\frac{45\,\delta}{2\pi^3 g_{*s}}
\sqrt{\frac{10}{g_*}}
\left(\frac{T_{\rm RH}}{M_{P}}\right)^3,
\qquad \mx \ll T_{\rm RH},
\end{equation}
where the entropy density is $s = \frac{2\pi^2 g_{*s} T^3}{45}$ and $g_{*s}$ denotes the relativistic degrees of freedom contributing to the entropy.

However, for DM heavier than the reheating temperature (\(M_{\rm X} \gg T_{\rm RH}\)), production occurs during the initial reheating phase driven by inflaton decay. As the Universe cools from its maximum temperature \(T_{\rm max}\), reached at the onset of reheating, down to temperatures \(T \sim M_{\rm X}\), DM can be efficiently produced during this stage. Assuming a matter-dominated reheating phase, the temperature and Hubble rate scale as \(T \propto a^{-3/8}\) and \(H \propto a^{-3/2}\), respectively.
\footnote{
Note that in the present analysis we focus on the PBH-dominated regime, i.e.\ \(\beta > \beta_{\rm c}\). In this case, the Universe undergoes multiple reheating phases: an initial reheating phase due to inflaton decay, followed by PBH-formation during radiation domination; subsequently, the Universe enters a PBH-dominated era and undergoes a final reheating phase driven by PBH evaporation. To avoid confusion, we denote the reheating temperature associated with inflaton decay by \(T_{\rm RH}\), while the reheating temperature resulting from PBH evaporation is denoted by \(T_{\rm ev}\).
}. We integrate Eq.~\eqref{eq:grav_BE} from the scale factor $a_{\rm max}$, corresponding to $T_{\rm max}$, to $a_{\rm X}$, defined by $T(a_{\rm X}) \simeq M_{\rm X}$. DM production beyond this point is neglected, since the thermal bath becomes kinematically incapable of producing particles heavier than the temperature. The resulting DM number density at the end of reheating is
\begin{equation}
n_{\rm X}(a_{\rm RH}) \simeq
\delta \sqrt{\frac{20}{g_*}}\,
\frac{T_{\rm RH}^6\,a_{\rm X}^{3/2}}{\pi M_{P}^3\,a_{\rm RH}^{3/2}}
\left[1-\left(\frac{a_{\rm max}}{a_{\rm X}}\right)^{3/2}\right].
\end{equation}
For $a_{\rm X} \gg a_{\rm max}$, this expression simplifies to
\begin{equation}
n_{\rm X}(a_{\rm RH}) =
\delta \sqrt{\frac{20}{g_*}}\,
\frac{T_{\rm RH}^{10}}{\pi M_{P}^3\,\mx^4}.
\end{equation}

As before, for $\beta < \beta_{\rm c}$, entropy is conserved after reheating, yielding
\begin{equation}
Y_0 =
\frac{45\,\delta}{2\pi^3 g_{*s}}
\sqrt{\frac{20}{g_*}}
\frac{T_{\rm RH}^7}{M_{P}^3\,\mx^4},
\qquad \mx \gg T_{\rm RH}.
\end{equation}

\paragraph{PBH-dominated scenario ($\beta > \beta_{\rm c}$):}
If PBHs dominate the energy density of the Universe prior to evaporation, their subsequent evaporation injects a substantial amount of entropy, thereby diluting the abundance of DM. Although the yield becomes constant again after evaporation, its value is reduced compared to the $\beta < \beta_{\rm c}$ case.

The yield of DM at evaporation can be related to its value at the early radiation--matter equality ($a_{\rm eq}$ ) as
\begin{equation}\label{Eq: betaconnection}
Y(a_{\rm ev}) \simeq Y(a_{\rm eq})\,
\frac{T_{\rm ev}}{T_{\rm eq}},
\end{equation}
During RD, $\rho_  R \propto a^{-4}$, while the PBH energy density scales as $\rho_{\rm BH} \propto a^{-3}$. For $\beta > \beta_{\rm c}$, PBH domination occurs at $
a_{\rm eq} = a_{\rm in}/\beta$. Since the radiation temperature scales as $T \propto a^{-1}$, the temperature at equality is $T_{\rm eq} = \beta\,T_{\rm in}$. The formation temperature is related to the initial PBH mass $M_{\rm in}$ via
\begin{equation} \label{Eq: Tin}
T_{\rm in} =
\sqrt{4\pi\gamma}
\left(\frac{90}{\pi^2 g_*}\right)^{1/4}
M_{P}\sqrt{\frac{M_{P}}{M_{\rm in}}}.
\end{equation}
Utilizing Eqns.\eqref{Eq: betaconnection},\eqref{Eq: Tin} and \eqref{eq:Teva}, the present-day yield in the PBH-dominated case is related to the non-dominated result as
\begin{equation}
Y_0(\beta > \beta_{\rm c}) \simeq
Y_0(\beta < \beta_{\rm c})\,
\sqrt{\frac{\epsilon}{2\pi\gamma}}\,
\frac{1}{\beta}
\left(\frac{M_{P}}{M_{\rm in}}\right).
\end{equation}

Finally, in both scenarios, the relic abundance of DM must reproduce the observed value $\Omega_{\rm X} h^2 \simeq 0.12$. This requirement relates the present-day yield to the DM mass via
\begin{equation}
M_{ \rm X} Y_0 =
\Omega_{\rm X} h^2 \frac{\rho_c}{s_0}
\simeq 4.3 \times 10^{-10}~\mathrm{GeV},
\end{equation}
where $s_0 \simeq 2.60 \times 10^3~\mathrm{cm^{-3}}$ is the present entropy density and $\rho_c \simeq 1.05 \times 10^{-5} h^2~\mathrm{GeV\,cm^{-3}}$ is the present-day critical density of the Universe~\cite{beringer2012particle}.

\begin{figure}[ht!]
    \centering
    \includegraphics[scale=0.32]{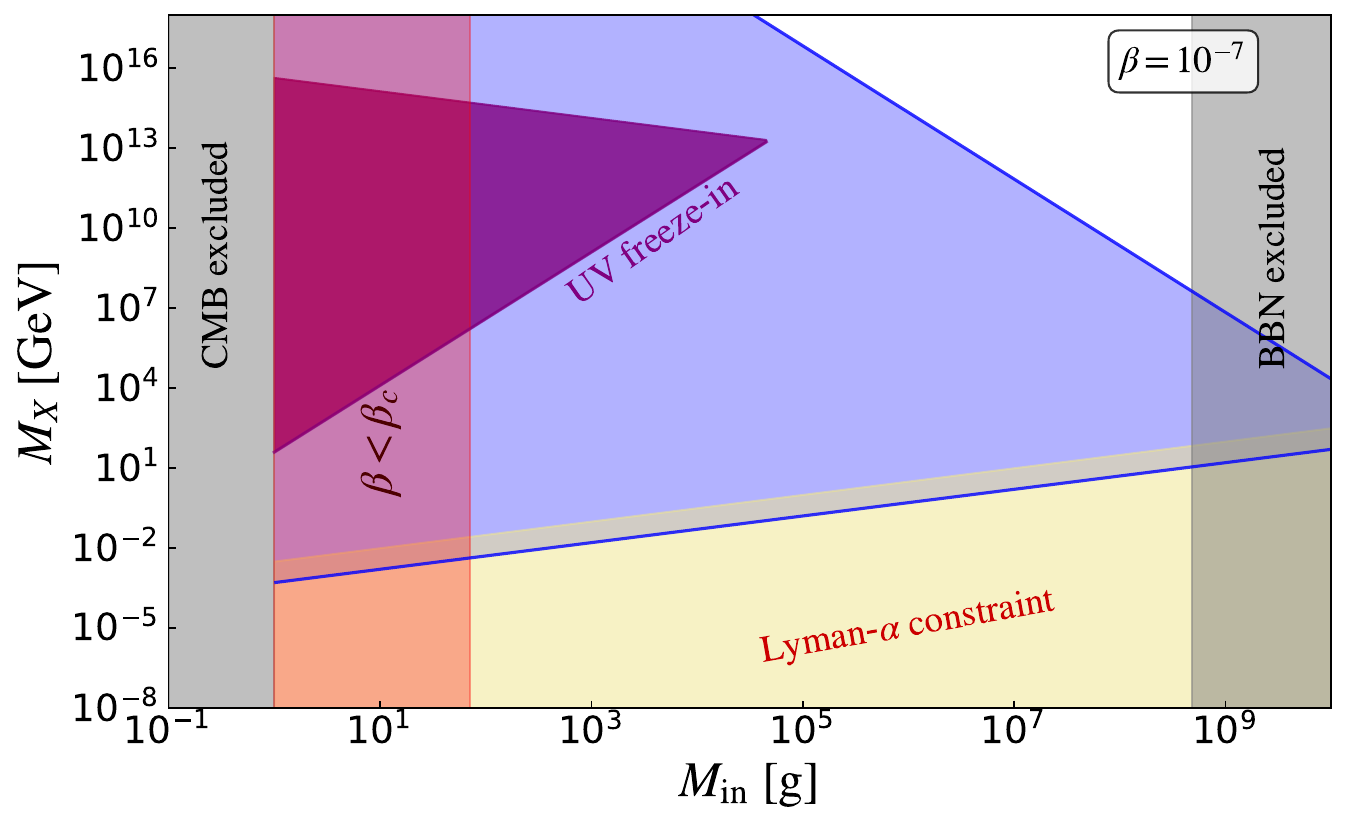}
    \includegraphics[scale=0.32]{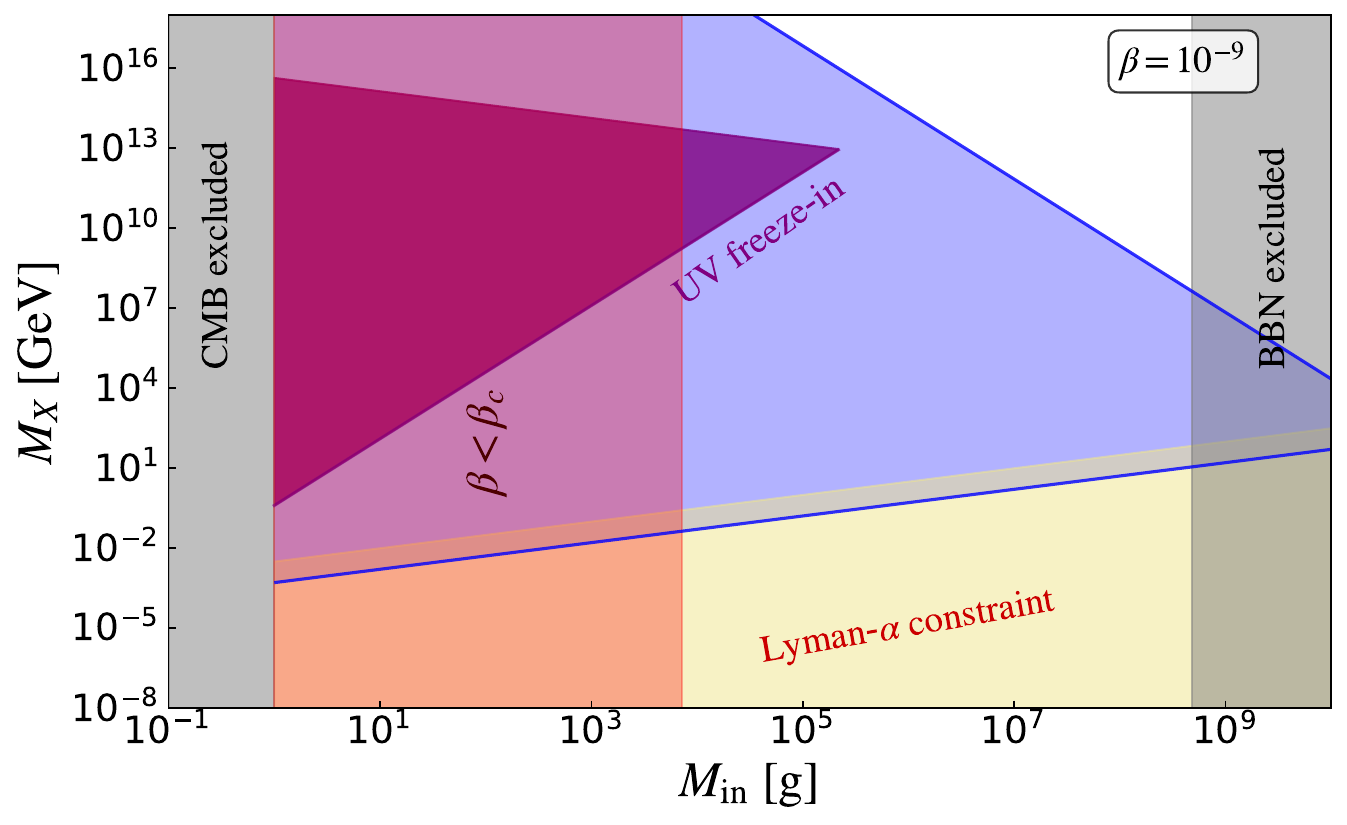}
    \caption{\it Allowed parameter range in the $M_{\rm in}$-$\mx$ plane, considering the gravitational production of DM from the SM thermal bath together with PBH-sourced DM. The purple shaded region corresponds to overproduction of DM due to gravitational freeze-in, while the blue-shade denotes overproduction from PBH evaporation. Regions excluded from Lyman-$\alpha$ observations are shown by light orange shade, and the grey-hatched areas are ruled out by BBN and CMB constraints. The \textbf{left panel} corresponds to $\beta=10^{-7}$, while the \textbf{right panel} is for $\beta=10^{-9}$. The red-hatched region indicates $\beta<\beta_c$, where the PBH never dominate the energy budget of the Universe.}
    \label{fig:grav_prod}
\end{figure}

Throughout our analysis, we set the reheating temperature, \textit{i.e.}\ $T_{\rm RH} = T_{\rm in}$, in order to ensure a scenario in which PBHs form during the RD era. The Fig.~\ref{fig:grav_prod} illustrates the allowed parameter range in the $M_{\rm in}$-$\mx$ plane, when gravitational production of DM from the thermal bath is taken into account alongside PBH evaporation, for two representative values of the initial PBH abundance, $\beta=10^{-7}$ and $10^{-9}$. For each value of \(\beta\), a triangular purple-shaded region appears in which gravitational freeze-in alone would overproduce DM. The boundary of this triangular region is determined by the requirement that the gravitationally produced DM satisfy the observed relic abundance. Superimposed on this region is the contribution from PBH-sourced DM production, shown in blue. We observe that, once DM production from  PBH evaporation is included, the gravitational contribution is subdominant compared to PBH evaporation for the scenarios of interest. Consequently, in the remainder of our analysis, we safely neglect gravitational production of DM and focus instead on PBH-sourced DM and annihilating DM for a given interaction cross section, rather than on gravitationally produced DM from the thermal bath.

\subsection{Freeze-in production of DM}
\label{sec:FI_GW}
If DM possesses extremely feeble interactions with the thermal bath, it never reach thermal equilibrium. In such a case, DM is gradually produced from the plasma via annihilation processes, a mechanism known as thermal \emph{freeze-in} (for review Ref.~\cite{Hall:2009bx}).

The presence of the non-gravitational interaction of DM with SM-bath, modified the Boltzmann equations, which are~\cite{Giudice:2000ex,Chung:1998rq}
\begin{eqnarray}
\label{eq:BEq}
\frac{d\rho_R}{da}+4\frac{\rho_R}{a}
&=&-\frac{\rho_{\rm BH}}{M_{\rm BH}}\frac{dM_{\rm BH}}{da} + 2\langle E_{\rm X} \rangle \frac{\langle \sigma v\rangle}{aH} \left(n_{\rm X}^2 - (\nxeq)^2\right) , \nonumber\\
\frac{d\rho_{\rm BH}}{da}+3\frac{\rho_{\rm BH}}{a}
&=&\frac{\rho_{\rm BH}}{M_{\rm BH}}\frac{dM_{\rm BH}}{da}, \nonumber\\
\frac{dn_{\rm X}}{da}+3\frac{n_{\rm X}}{a}
&=&\Gamma_{\rm BH\to X}\frac{\rho_{\rm BH}}{M_{\rm BH}aH} - \frac{\langle \sigma v\rangle}{aH} \left(n_{\rm X}^2 - (\nxeq)^2\right), \nonumber\\
\frac{dM_{\rm BH}}{da}&=&-\epsilon\frac{\Mp^4}{M_{\rm BH}^2aH}.
\end{eqnarray}
Here, $\langle \sigma v\rangle$ is the thermally averaged interaction cross-section of the DM and $\langle E_{\rm X}\rangle\simeq\sqrt{\mx^2 + 9T^2}$ is the average energy per DM particle. The equilibrium number density of the DM, denoted by $\nxeq$, can be expressed as~\cite{Giudice:2000ex}
\begin{eqnarray}
\label{eq:nx_eq}
    \nxeq = \frac{g_{X}T^3}{2\pi^2}\left(\frac{\mx}{T}\right)^2K_2\left(\frac{\mx}{T}\right),
\end{eqnarray}
where $K_2$ is the modified Bessel function of second kind. The set of equations (Eqs.~\eqref{eq:BEq} and \eqref{eq:nx_eq}) is solved numerically to obtain the yield of DM at late times, when there is no further DM production and no additional entropy injection in the system. This yield remains constant thereafter and corresponds to the present-day value, \(Y_0\).  

Freeze-in production is dominated near the highest temperature of the thermal bath. Since the annihilation cross-section is too small to reach the thermal equilibrium, the number density of DM always satisfies $\nx\ll\nxeq$. Thus, the inverse processes can be safely neglected. This results in the relic abundance of DM, produced by freeze-in mechanism, increases with the increasing $\langle \sigma v\rangle$ for a particular DM mass. The PBH-domination injects the entropy in the Universe, leading to the dilution of the relic abundance of DM. Thus, PBH-domination allows comparatively larger $\langle \sigma v\rangle$ for the freeze-in mechanism in order to satisfy the correct relic. Incorporating the three production mechanisms, PBH-sourced DM, gravitational and thermal freeze-in production of DM, the present-day DM relic abundance is given by the sum of individual contributions,
\begin{equation}
\label{eq:Omega_total}
\Omega_X h^2 =
\Omega_X^{\rm PBH} h^2 + \Omega_X^{\rm FI} h^2 + \Omega_X^{\rm grav} h^2,
\end{equation}
although, as emphasized earlier, the gravitational contribution
$\Omega_X^{\rm grav} h^2$ is negligible compared the other two.

\begin{figure}[ht!]
    \centering
    \includegraphics[scale=0.30]{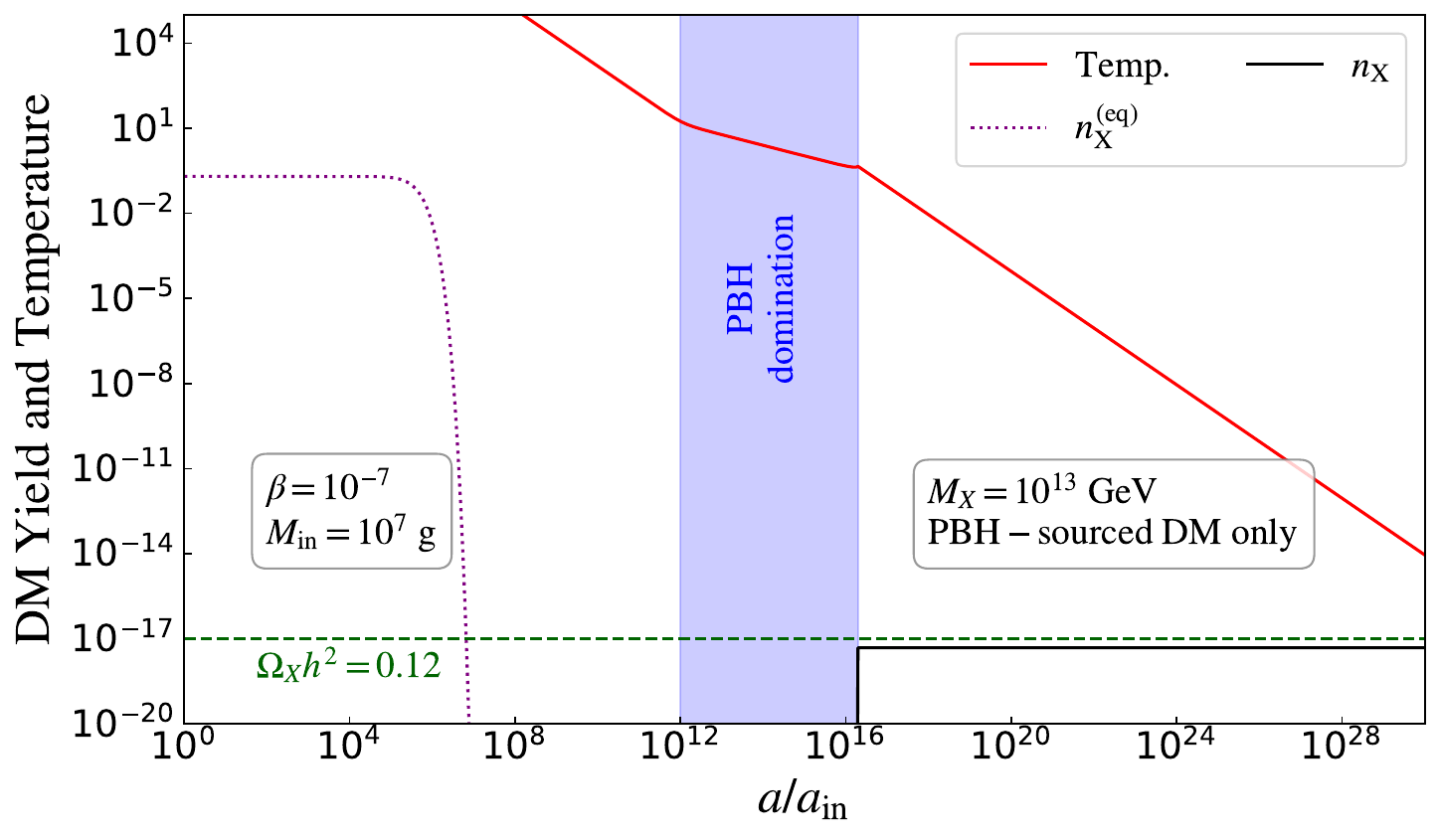}
    \includegraphics[scale=0.30]{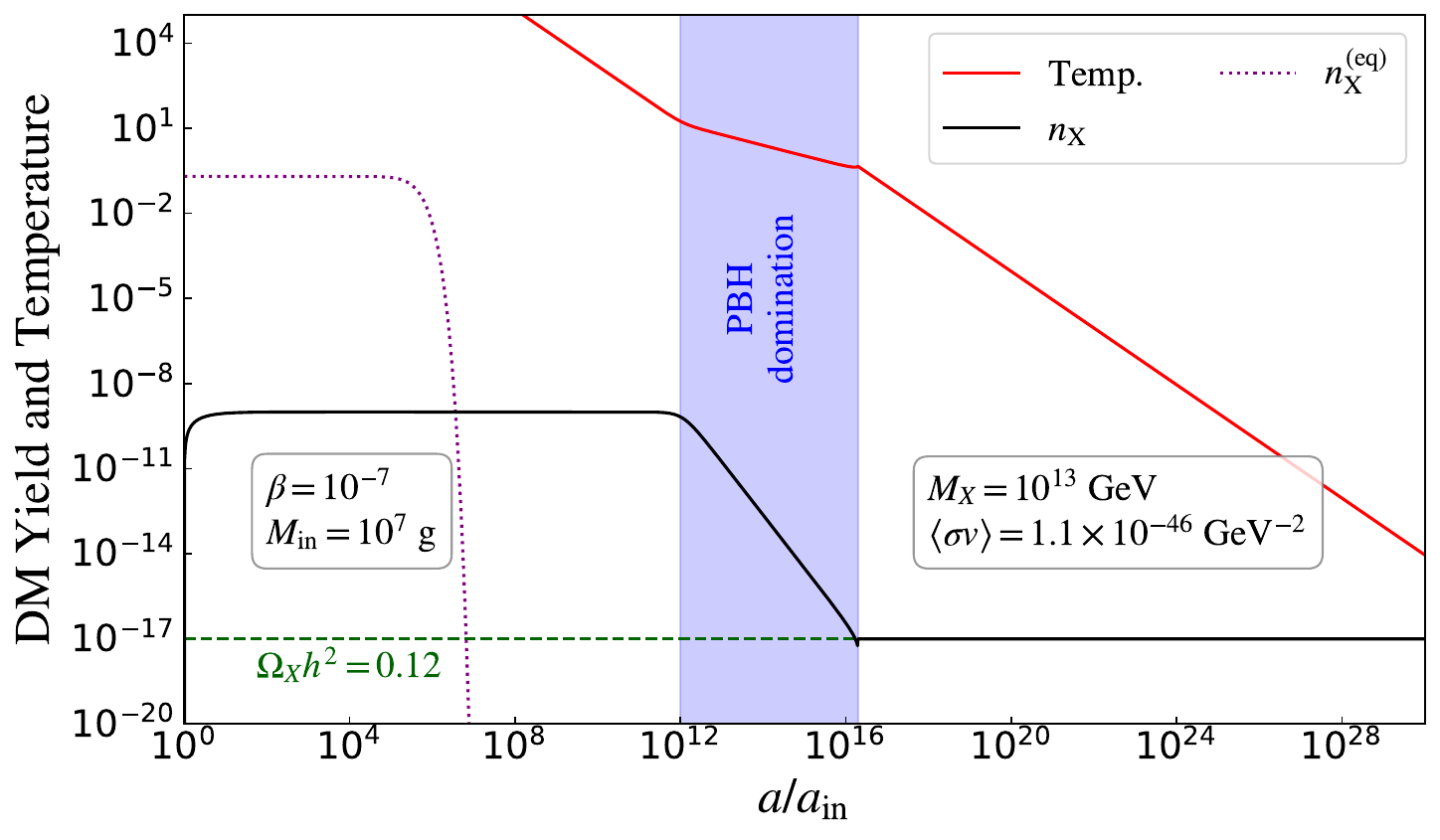}
    \caption{\it Evolution of yield of DM and temperature for PBH reheating scenario. The red curve shows the evolution of temperature. Each panel displays the yield of DM (black solid line) and equilibrium number density (purple dotted line), for $\mx=10^{13}$ GeV, $M_{\rm in}=10^7$ g and $\beta=10^{-7}$. The \textbf{left panel} corresponds to PBH-sourced DM production. The \textbf{right panel} shows the freeze-in production, with $\langle \sigma v\rangle=1.1\times 10^{-46}$ GeV$^{-2}$, in addition to PBH-sourced production. } 
    \label{fig:DMyeild_freezein}
\end{figure} 

\begin{figure}[ht!]
    \centering
    \includegraphics[scale=0.30]{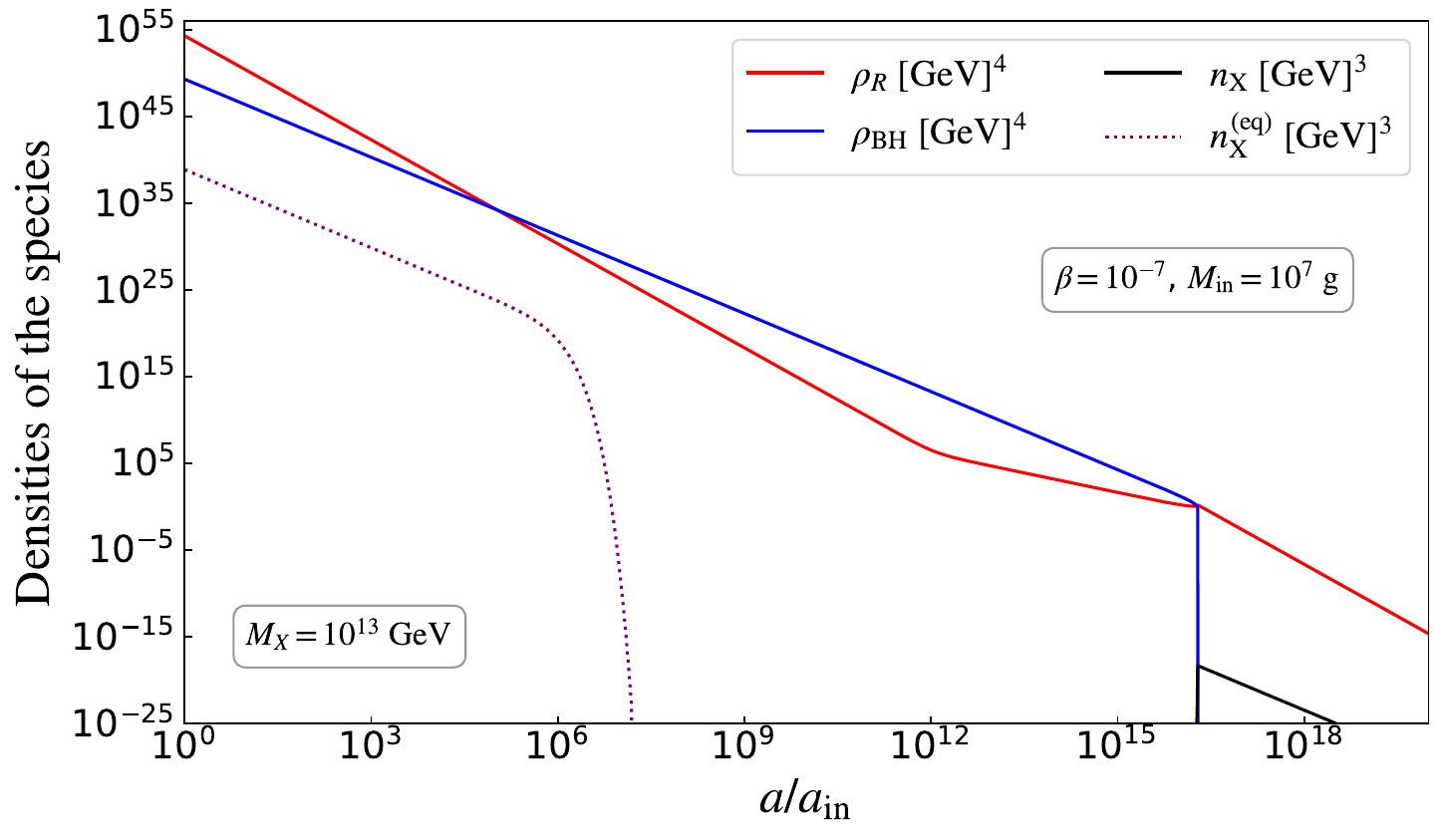}
    \includegraphics[scale=0.30]{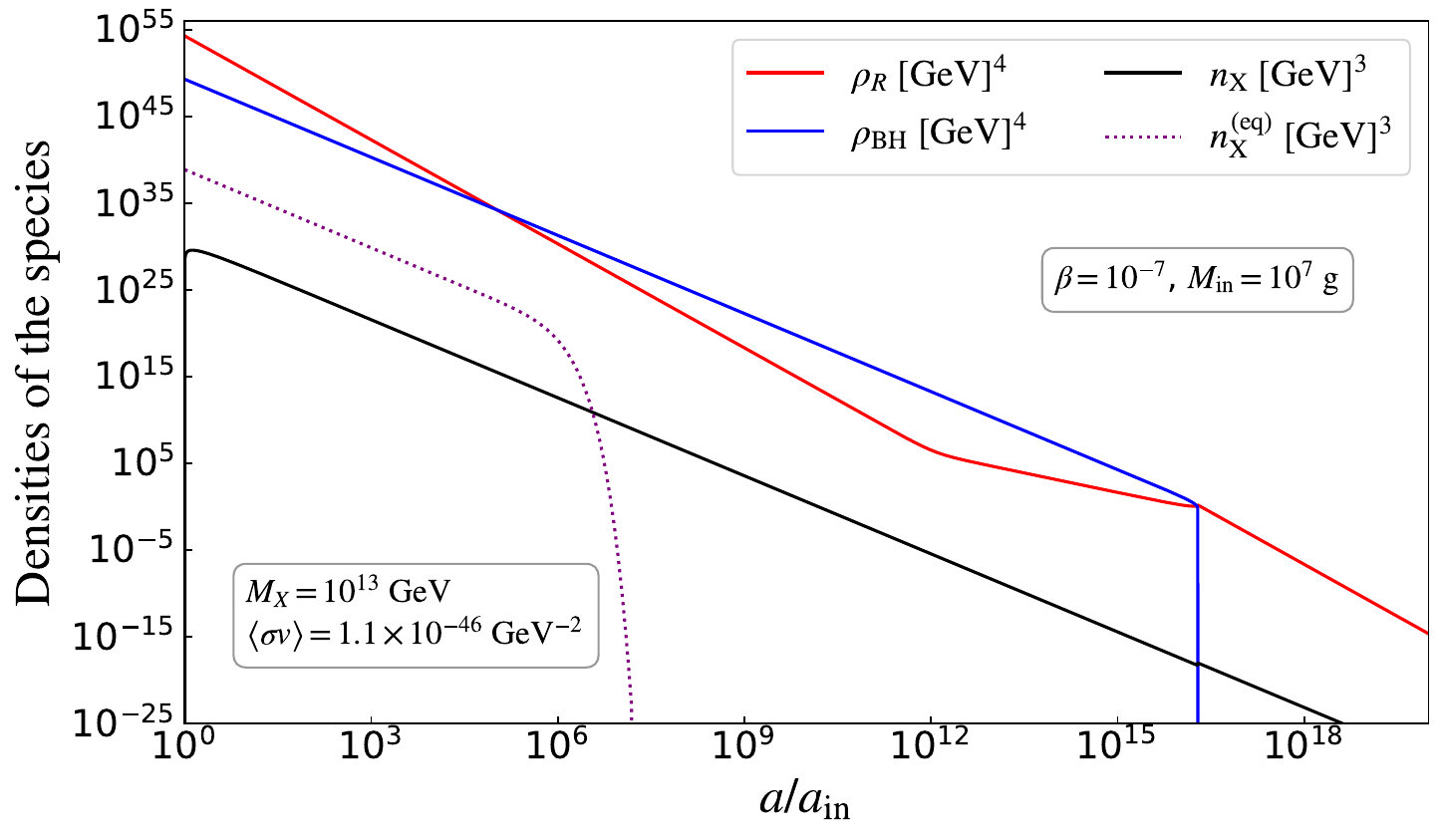}
    \caption{\it Evolution of radiation, PBH and DM density. The red and blue curves show the evolution of energy densities of radiation and PBH, respectively, whereas the black line is for the number density of DM, all plotted as a function of $a/a_{\rm in}$, obtained by numerically solving Eq.~\eqref{eq:BEq}. The dotted curve in each panel denotes the equilibrium number density of DM for $M_X=10^{13}$ GeV and $M_{\rm in}=10^7$ g. In the \textbf{left panel} DM is solely produced via PBH evaporation and remains under-abundant. In the \textbf{right panel} the inclusion of freeze-in production, with $\langle \sigma v\rangle=1.1\times 10^{-46}$ GeV$^{-2}$, in addition to PBH-sourced production yields the observed DM relic abundance.}
    \label{fig:evolution_freezein}
\end{figure}
In In Figs.~\ref{fig:DMyeild_freezein} and \ref{fig:evolution_freezein}, we illustrate the evolution of the DM yield and energy density, respectively when production occurs solely through PBH evaporation, as shown in the left panel, and when an additional production channel, namely DM annihilation via the freeze-in mechanism is included, as shown in the right panel. For a fixed choice of the PBH and DM masses, $M_{\rm in}=10^{7}\,\mathrm{g}$ and $M_{X}=10^{13}\,\mathrm{GeV}$ with $\beta=10^{-7}$, it is evident from the left panel that DM production from PBH evaporation alone is insufficient to account for the observed relic abundance, resulting in an underabundant region. However, once freeze-in production from the thermal bath is taken into account, with an annihilation cross section $\langle\sigma v\rangle =1.1\times 10^{-46}\,\mathrm{GeV^{-2}}$, the correct DM relic abundance can be achieved. This is clearly demonstrated in the right panel of Fig.~\ref{fig:DMyeild_freezein}, where the additional freeze-in contribution compensates for the deficit arising from PBH evaporation alone.

\begin{figure}[ht!]
    \centering
    \includegraphics[scale=0.31]{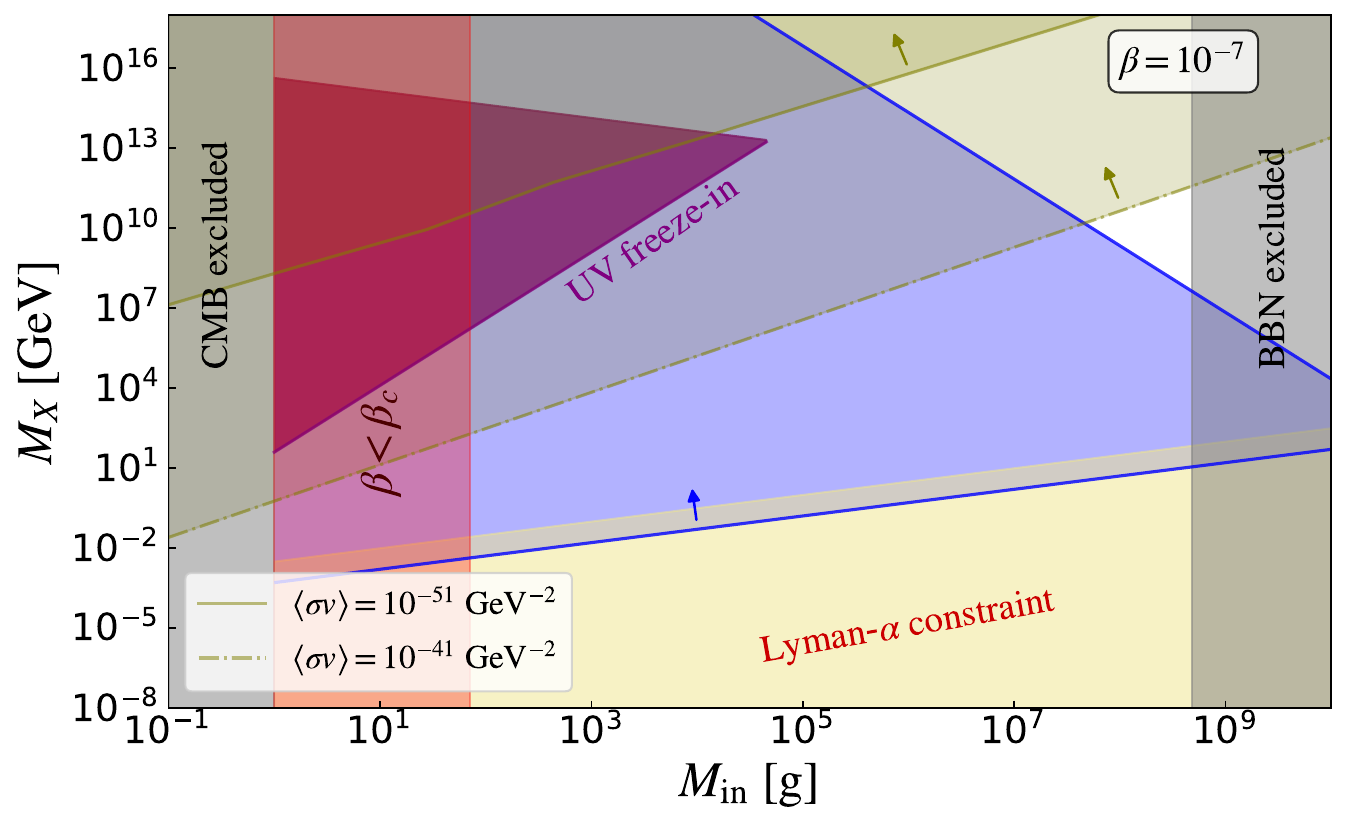}
    \includegraphics[scale=0.31]{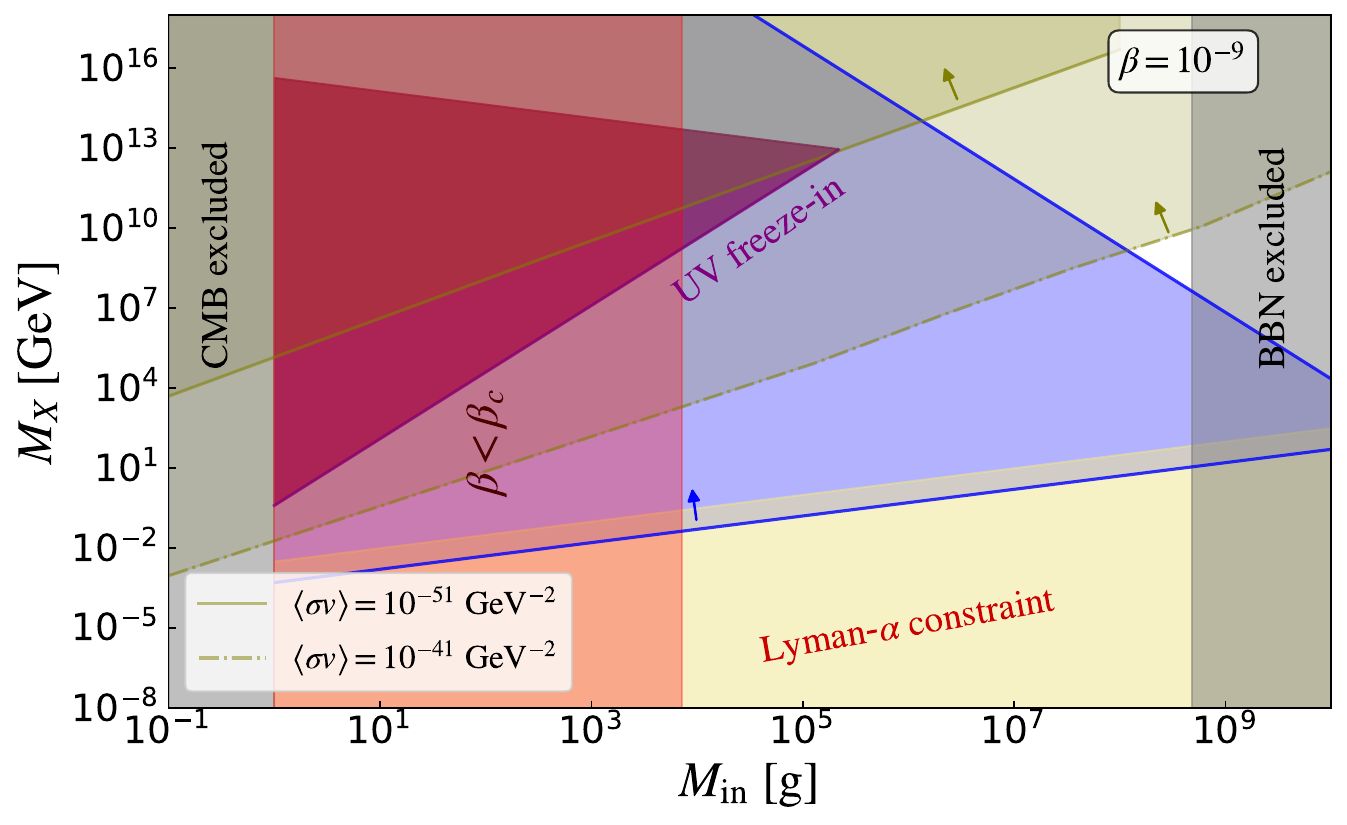}
    \caption{\it Allowed parameter range in the $M_{\rm in}$-$\mx$ plane, incorporating the production of DM from PBH in addition to freeze-in production of DM from thermal bath. \textbf{Left plot} is for $\beta =10^{-7}$, while \textbf{right plot} presents $\beta=10^{-9}$. The blue-shaded region, in both the panels, is excluded due to the overproduction of DM, while the white regions are allowed, without freeze-in production of DM. The arrows indicate where $\Omega_{\rm X}h^2>0.12$. The red-hatched region indicates the range of PBH-mass which lies inside $\beta<\beta_c$. Including freeze-in production further shrinks the allowed region, shown by the olive shading. A larger interaction cross-section reduces the allowed region even more. The grey-hatched areas indicate the regions excluded by BBN and CMB constraints.}
    \label{fig:MBH_MX}
\end{figure}

Fig.~\ref{fig:MBH_MX} summarizes the combined effects of all three DM production mechanisms for two representative values of the initial PBH fraction, $\beta = (10^{-7},\,10^{-9})$. Since our analysis focuses on the PBH-dominated regime, the region $\beta < \beta_c$ is explicitly shaded with red hatches. When only PBH-sourced DM production is taken into account, the DM relic abundance remains underabundant for sufficiently heavy DM masses, namely $\mx \in (10^{7},\,10^{18})\,\mathrm{GeV}$, over a wide range of PBH-formation masses $M_{\rm in} \in (10^{8},\,10^{4})\,\mathrm{g}$. The inclusion of thermal freeze-in production allows these underabundant regions to reach the observed DM relic abundance. However, once the annihilation cross section is fixed to sufficiently large values, freeze-in production can also overproduce DM, thereby reducing the viable parameter space in the $\mx$--$M_{\rm in}$ plane. These excluded regions, characterized by $\Omega_X h^2 > 0.12$, are indicated by arrows in Fig.~\ref{fig:MBH_MX}. The dependence of the annihilation cross section $\langle\sigma v\rangle$ required to satisfy the correct relic abundance on $\beta$ can be understood as follows. Smaller values of $\beta$ correspond to a shorter duration of PBH domination, leading to a reduced period of entropy injection from PBH evaporation. Consequently, comparatively smaller values of $\langle\sigma v\rangle$ are sufficient to reproduce the observed relic abundance. For a fixed DM mass and PBH-formation mass $M_{\rm in}$, increasing $\beta$ therefore requires a larger value of $\langle\sigma v\rangle$ to satisfy the correct relic. This behavior is clearly visible when comparing the left and right panels of Fig.~\ref{fig:MBH_MX}. For instance, the overproduction region corresponding to $\langle\sigma v\rangle = 10^{-41}\,\mathrm{GeV}^{-2}$ is significantly larger for $\beta = 10^{-9}$ than for $\beta = 10^{-7}$.

Overall, the combined effect of freeze-in and PBH evaporation significantly constrains the  parameter space. While PBH evaporation dominates DM production for light masses, freeze-in becomes increasingly important for heavier DM, especially in regions that would otherwise remain underabundant.

\subsection{Freeze-out production of DM }
\label{sec:FO_GW}
Let us now turn our attention to another  production mechanism  of DM, namely, the freeze-out production, and discuss how it further modifies the above scenario.
If the interaction strength between DM and SM bath is sufficiently large, DM can reach thermal equilibrium with SM the bath at the early time and freezes out once the annihilation rate becomes smaller than the Hubble expansion rate. This production mechanism is called \emph{freeze-out} mechanism~\cite{chiu1966symmetry,Kamionkowski:1990ni}. We use the set of Boltzmann equations Eq.~\eqref{eq:BEq} to obtain the comoving number density of  DM. For the freeze-out mechanism the relic abundance of  DM is inversely proportional to annihilation cross-section, unlike the freeze-in scenario.

\begin{figure}
    \centering
    \includegraphics[scale=0.31]{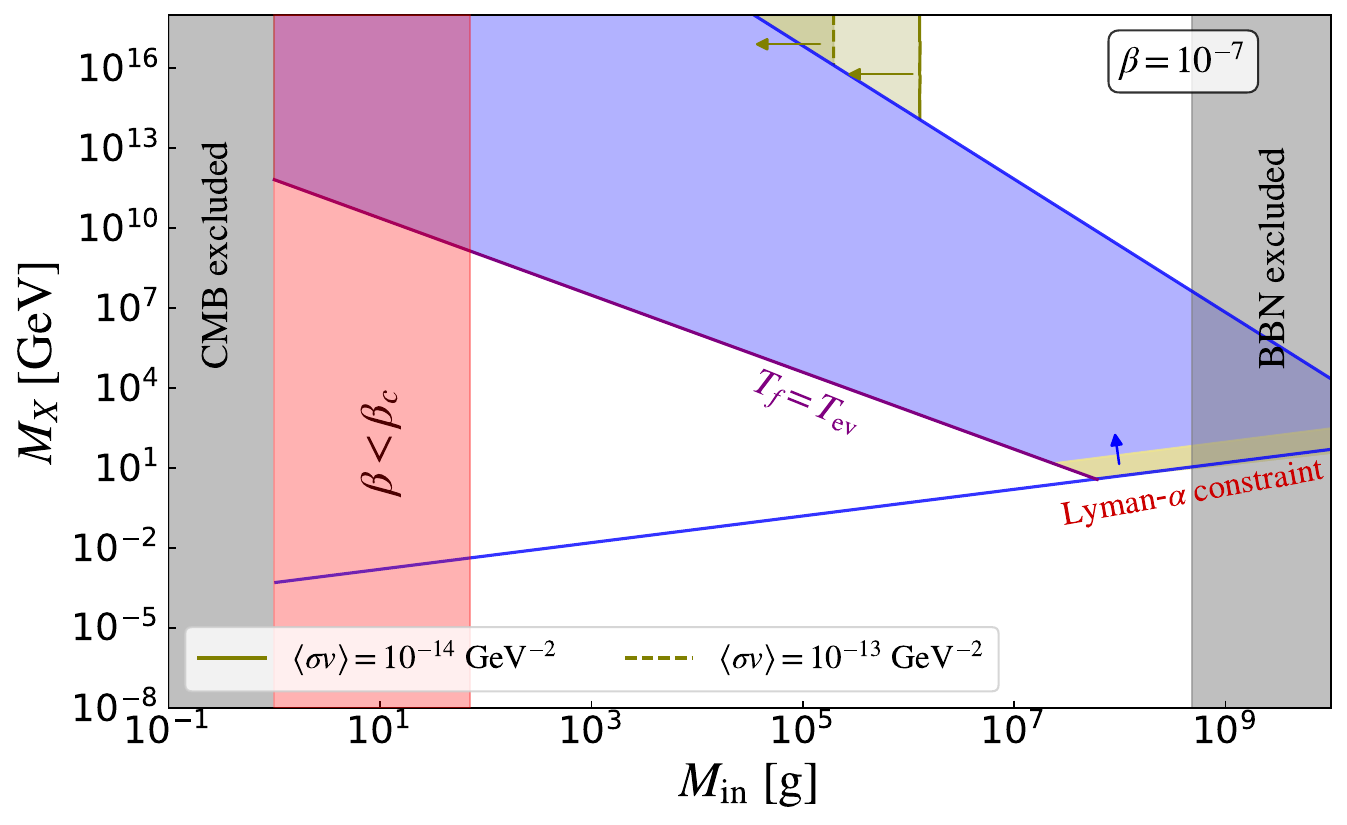}
    \includegraphics[scale=0.31]{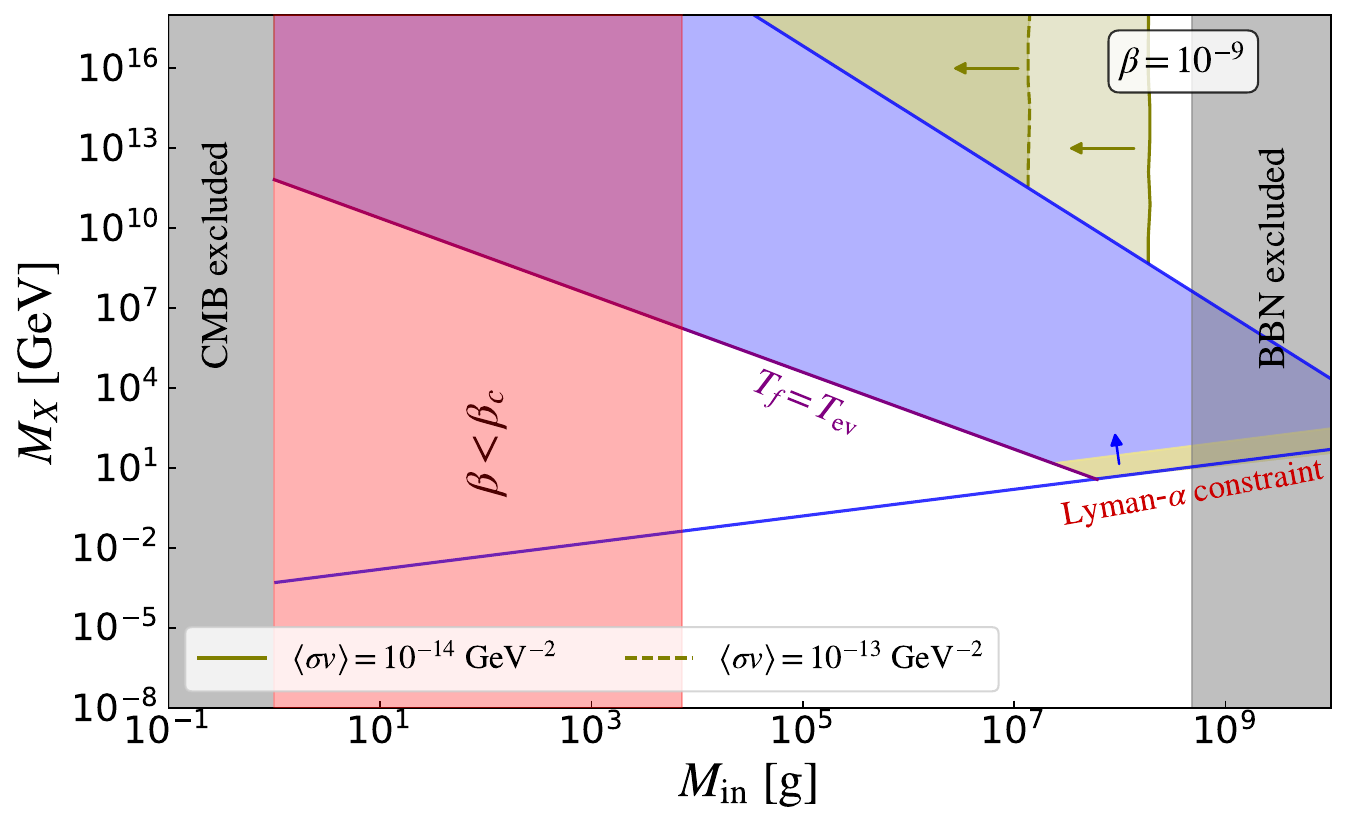}
    \caption{\it Allowed parameter space in the $M_{\rm in}$--$\mx$ plane including DM production from PBH evaporation together with freeze-out production. The blue-shaded region is excluded due to DM overproduction, while the white region remains allowed. The arrows indicate regions where $\Omega_{\rm X}h^2>0.12$. The solid and dashed olive curves denote the $\Omega_{\rm X}h^2=0.12$ contours for two representative values of DM annihilation cross-section, incorporating both production mechanisms. Freeze-out production partially reopens regions of parameter space that are excluded when only PBH-sourced DM is considered.}
    \label{fig:MBH_MX_fo}
\end{figure}

One interesting outcome of this analysis is the reopening of parameter
space in the $M_{\rm X}$–$M_{\rm in}$ plane that would otherwise be excluded
due to DM overproduction from evaporating PBHs. If DM freeze-out occurs after the completion of PBH evaporation,
$T_{\rm f} < T_{\rm ev}$, the DM particles produced during evaporation
thermalize with the plasma. Since freeze-out then takes place during the
standard RD era following PBH domination, the relic
abundance is determined by the usual thermal freeze-out mechanism.
Consequently, the correct DM relic density can be obtained for the standard
WIMP annihilation cross-section,
$\langle \sigma v \rangle \sim 10^{-9}\,{\rm GeV^{-2}}$.
This behaviour is clearly illustrated in Fig.~\ref{fig:MBH_MX_fo}, where a new
underabundant region appears below the line $T_{\rm f}=T_{\rm ev}$. In contrast, if freeze-out occurs before PBH evaporation is complete,
$T_{\rm f} > T_{\rm ev}$, the situation changes qualitatively. Since the
dominant DM production from PBH evaporation occurs near the final stages
of evaporation i.e., after freeze-out, the region that is overabundant
when considering only PBH production remains overabundant. However, the
region that would be underabundant from PBH production alone can now be
partially compensated by thermal production through the freeze-out
mechanism. In this case, the required annihilation cross-section must be modified to
reproduce the observed relic abundance. Because the DM mass is typically
much larger than the evaporation temperature, entropy injection from PBH
evaporation dilutes the DM yield. To obtain the correct relic density,
freeze-out must occur earlier, which requires a smaller annihilation
cross-section. Otherwise, the relic abundance would remain underabundant.
Nevertheless, a significant portion of this parameter space is excluded
by the unitarity bound, since the required cross-section violates
$\langle \sigma v \rangle \leq 8\pi/M_{\rm X}^2$ ~\cite{Giudice:2000ex}. 
\begin{figure}
    \centering
    \includegraphics[scale=0.30]{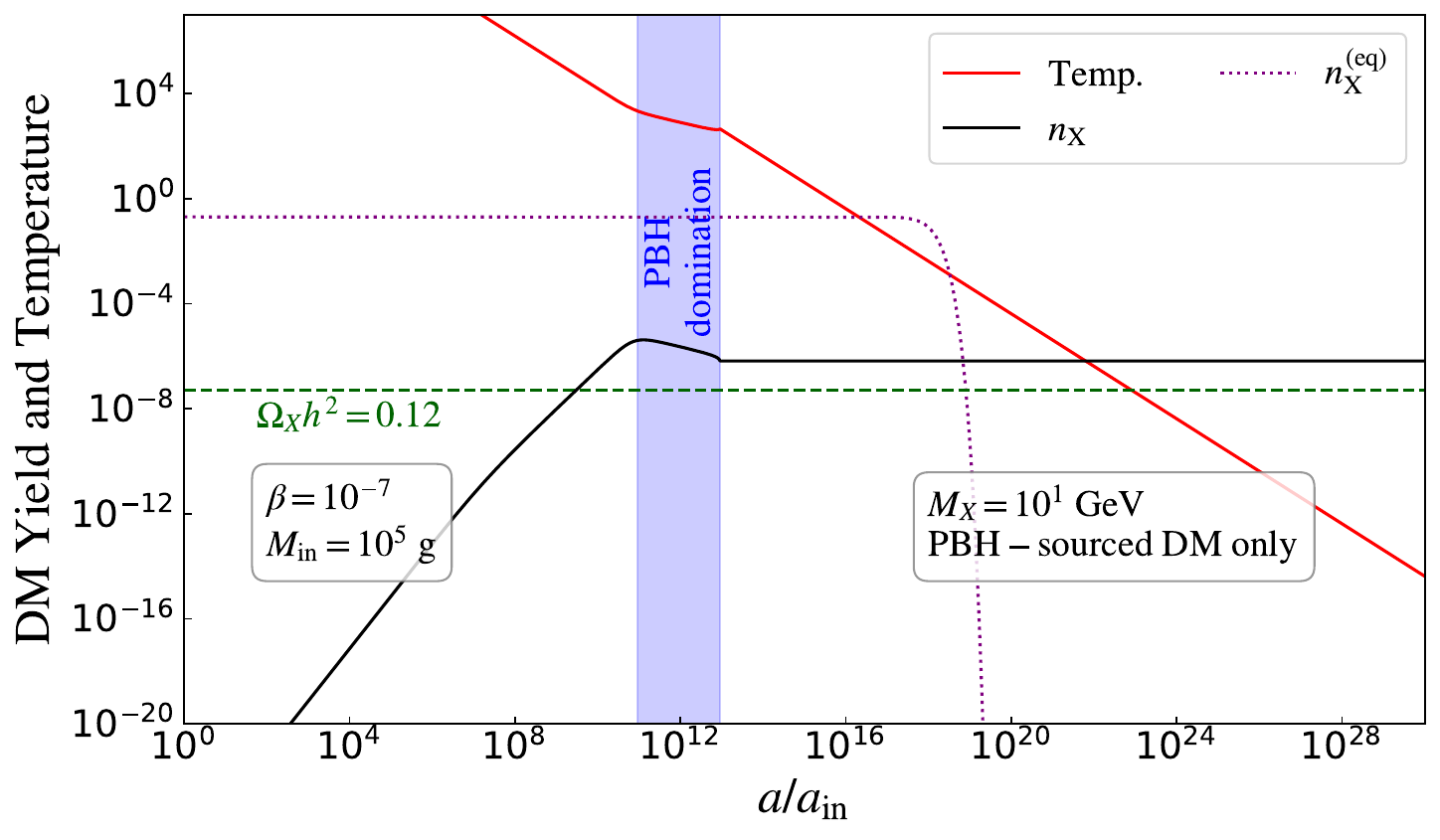}
    \includegraphics[scale=0.30]{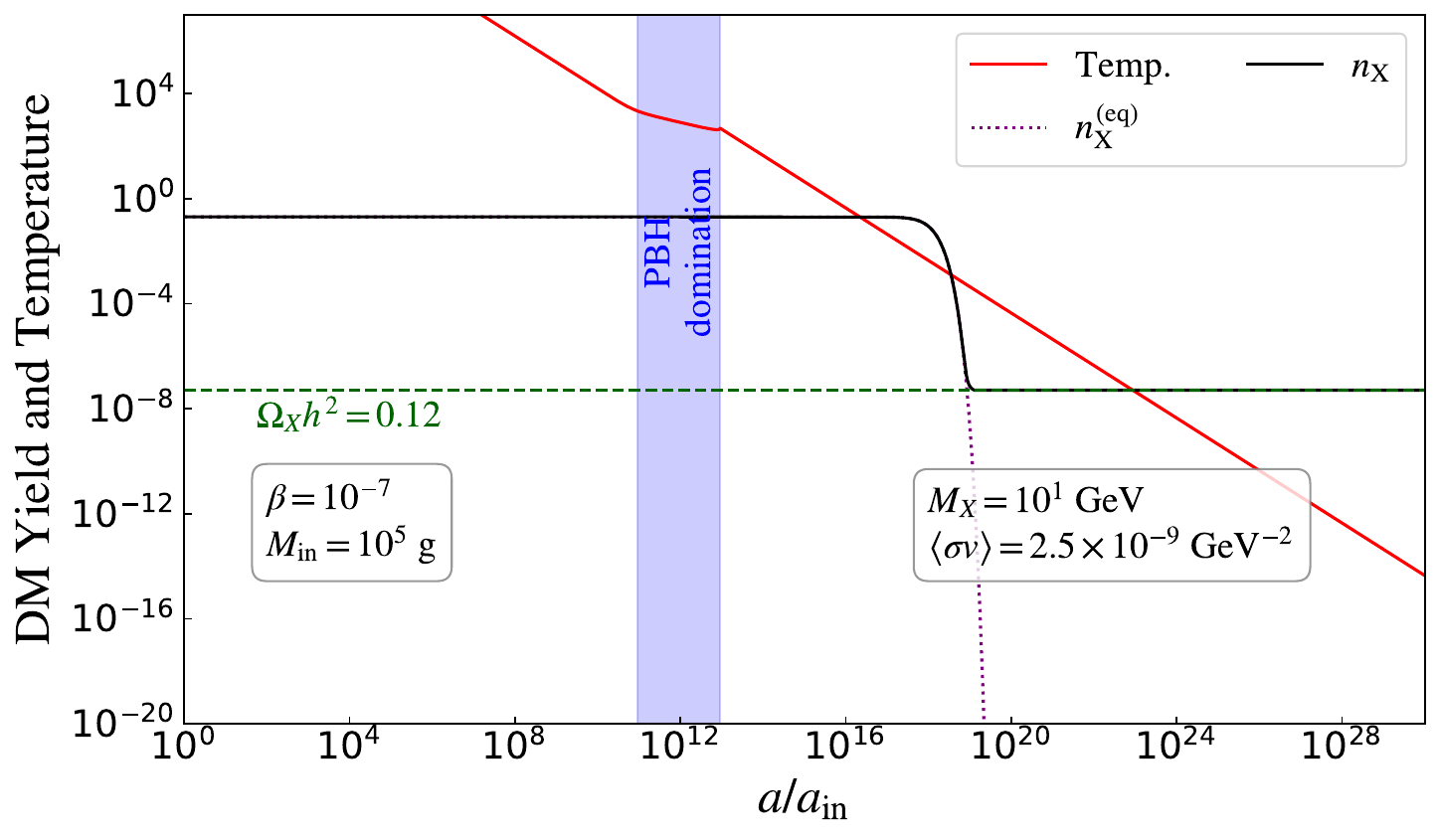}
    \caption{\it Evolution of yield of DM and temperature. The red curve shows the evolution of temperature. Each panel displays the yield of DM (black solid line) and equilibrium number density (purple dotted line), for $\mx=10^{1}$ GeV and $M_{\rm in}=10^5$ g. The \textbf{left panel} corresponds to PBH-sourced DM production which is over-abundant. The \textbf{right panel} shows the freeze-in production, with $\langle \sigma v\rangle=2.5\times 10^{-9}$ GeV$^{-2}$, in addition to PBH-sourced production, yields the observed DM relic abundance.}
    \label{fig:DMyeild_freezeout}
\end{figure}

\begin{figure*}
    \centering
    \includegraphics[scale=0.30]{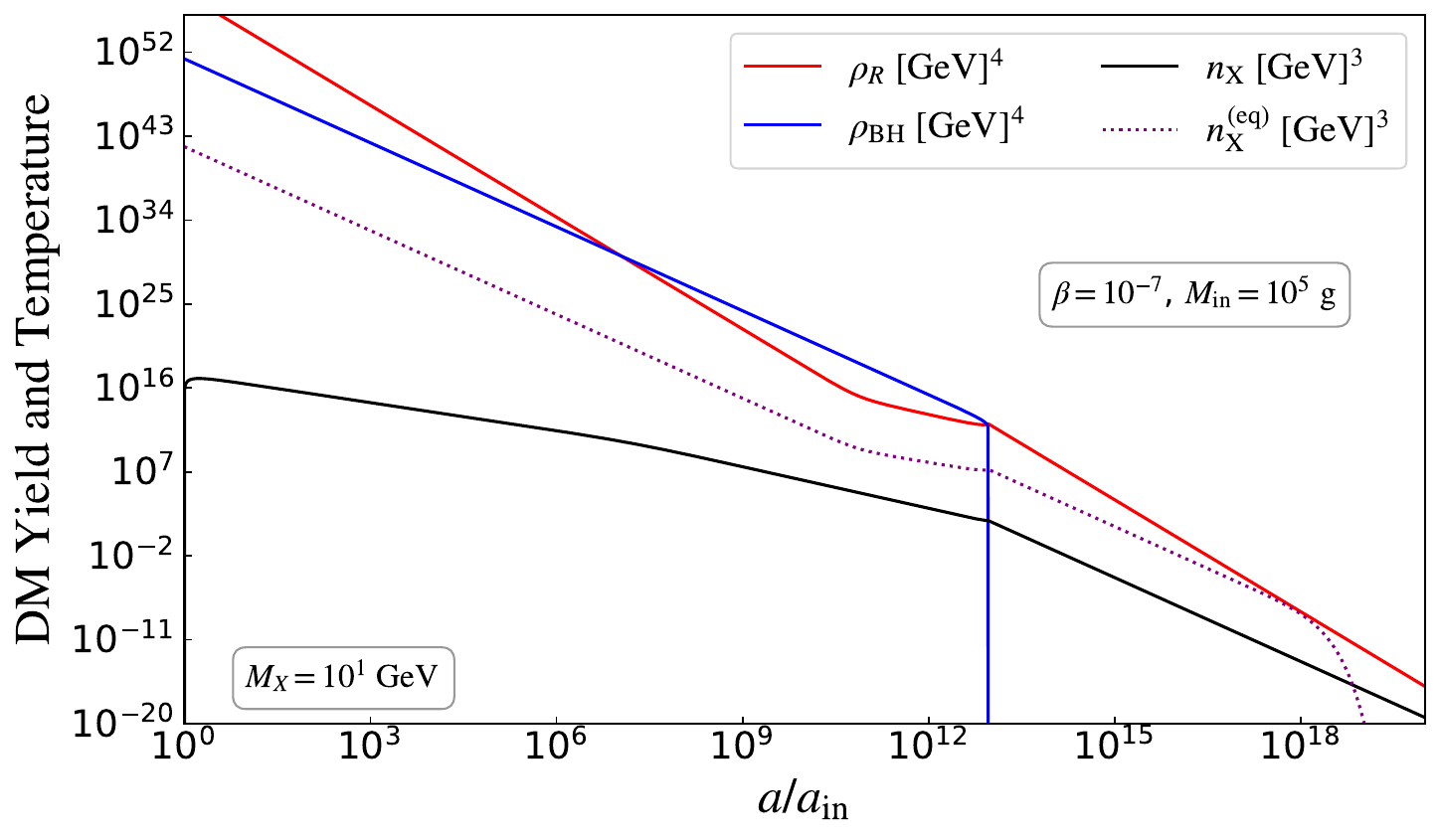}
    \includegraphics[scale=0.30]{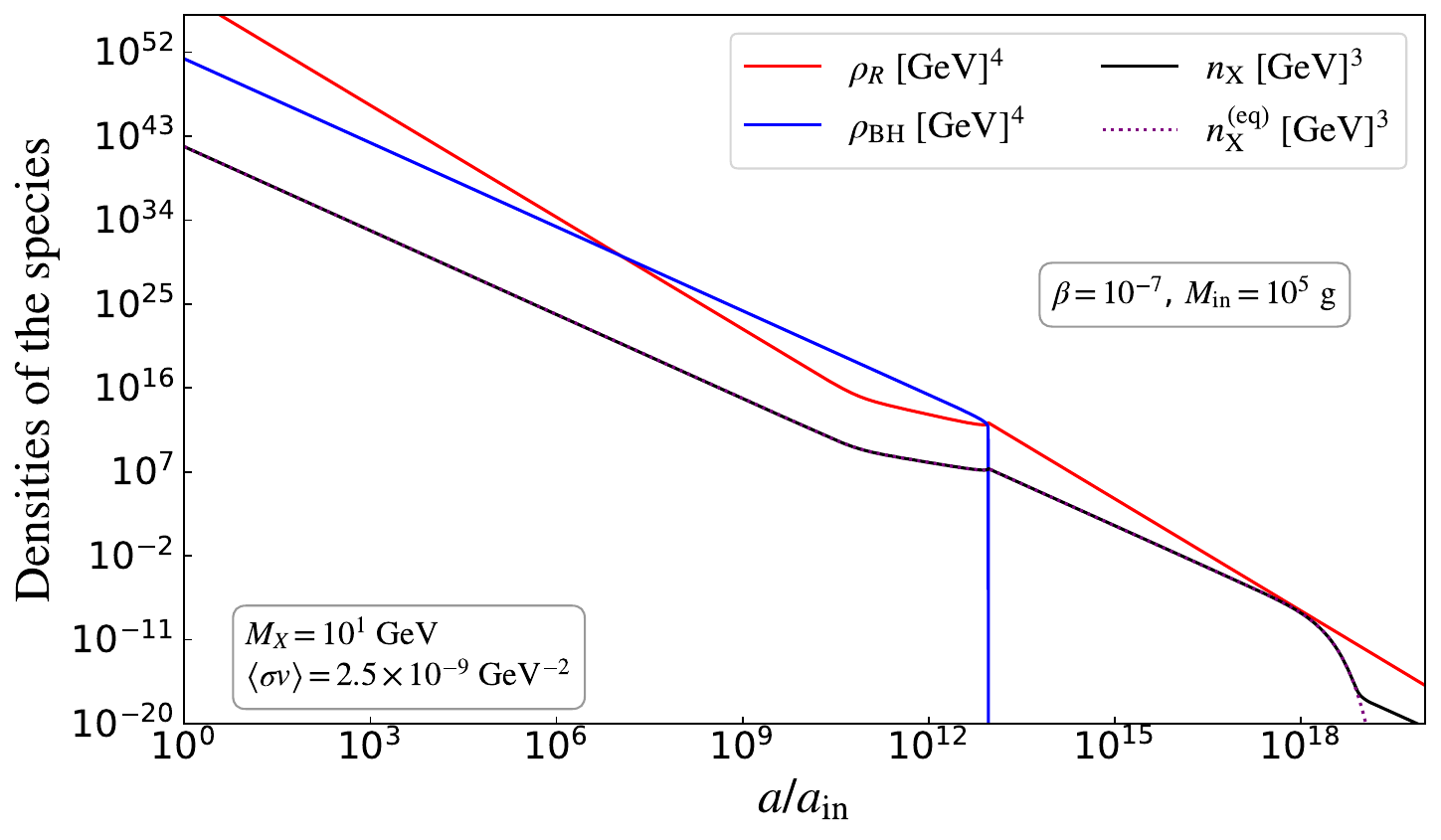}
    \caption{\it Evolution of radiation, PBH and DM density. The red and blue curves show the evolution of energy densities of radiation and PBH, respectively, whereas the black line is for the number density of DM, all plotted as a function of $a/a_{\rm in}$, obtained by numerically solving Eq.~\eqref{eq:BEq}. The dotted curve in each panel denotes the equilibrium number density of DM for $M_X=10^{1}$ GeV and $M_{\rm in}=10^5$ g. In the \textbf{left panel} DM is solely produced via PBH evaporation and is over-abundant. In the \textbf{right panel} the inclusion of freeze-out production, with $\langle \sigma v\rangle=2.5\times 10^{-9}$ GeV$^{-2}$, in addition to PBH-sourced production yields the observed DM relic abundance.}
    \label{fig:evolution_freezeout}
\end{figure*}

In Figs.~\ref{fig:DMyeild_freezeout} and \ref{fig:evolution_freezeout},
we show the evolution of the DM yield and energy density, respectively.
The left panels correspond to the case where DM production occurs solely
through PBH evaporation, while the right panels include an additional
production channel from DM annihilation via the thermal freeze-out
mechanism. For a representative choice of parameters,
$M_{\rm in}=10^{5}\,\mathrm{g}$,
$M_{\rm X}=10\,\mathrm{GeV}$, and
$\beta=10^{-7}$,
the left panel of Fig.~\ref{fig:DMyeild_freezeout} clearly shows that
PBH evaporation alone leads to an overproduction of DM.
However, once thermal freeze-out is included, the situation changes.
Since freeze-out occurs during the radiation-dominated era after PBH
evaporation, the DM particles thermalize with the plasma and their relic
abundance is determined by the standard freeze-out mechanism.
For the canonical WIMP annihilation cross-section,
$\langle\sigma v\rangle = 2.5 \times 10^{-9}\,\mathrm{GeV^{-2}}$,
the observed DM relic abundance can be successfully reproduced. This behaviour is clearly demonstrated in the right panel of
Fig.~\ref{fig:DMyeild_freezeout}, where the additional thermal
production channel regulates the DM abundance. In particular, DM
particles reach thermal equilibrium and remain in equilibrium for a
sufficiently long period before freezing out, thereby yielding the
correct relic density.

Having established the viable DM parameter space arising from the
various production mechanisms discussed above, we now turn to the
corresponding observational implications. While certain regions of
this parameter space may be probed through conventional indirect
detection searches, an early PBH-dominated epoch can also leave
distinct imprints in the gravitational-wave sector.
In particular, a period of PBH domination can source induced
gravitational waves, providing a complementary and independent probe
of the DM scenarios considered in this work. We explore this
connection in the following sections.

\section{Induced gravitational waves in PBH-induced reheating}
\label{sec:GW_PBH}

As discussed earlier, PBH reheating gives rise to two distinct and well-motivated sources of stochastic GWs: those induced by isocurvature fluctuations associated with the discrete PBH distribution, and those sourced by adiabatic curvature perturbations amplified during the sudden transition from PBH domination to radiation domination \footnote{Note that PBH formation is accompanied by large scalar curvature perturbations that source secondary GWs \cite{Baumann:2007zm, Espinosa:2018eve, Domenech:2019quo, Inomata:2023zup, Franciolini:2023pbf, Firouzjahi:2023lzg, Maity:2024odg}, while Hawking evaporation also generates a stochastic GW background through graviton emission \cite{Anantua:2008am,Dong:2015yjs,Ireland:2023avg}. For the ultralight PBHs studied here, both contributions are expected to lie at frequencies well above the sensitivity of existing and planned GW detectors. Consequently, we restrict our analysis to GWs sourced by PBH isocurvature fluctuations and the enhanced adiabatic perturbations during the PBH-dominated epoch.}. Without going to details, here we briefly summarize the theoretical framework and key results relevant for both contributions (see, for instance Refs.~\cite{Papanikolaou:2020qtd,Domenech:2020ssp,Dalianis:2020gup,Domenech:2021ztg,Domenech:2021wkk,Papanikolaou:2021uhe,Bhaumik:2022pil,Bhaumik:2022zdd,Papanikolaou:2022chm,Bhaumik:2024qzd,Domenech:2024wao,Gross:2024wkl,Gross:2025hia,Inomata:2020lmk,White:2021hwi,Bhaumik:2022pil,Bhaumik:2022zdd,Bhaumik:2023wmw,Bhaumik:2024qzd,Domenech:2024wao}). The GW spectral energy density, as a function of wavenumber ($k$), at any conformal time, $\eta$ is defined in terms of the oscillation-averaged tensor power spectrum as
\begin{eqnarray}
\Omega_{\rm GW}(k,\eta)=\frac{k^2}{12\mathcal{H}^2}\,
\overline{\mathcal{P}_h(k,\eta)}\,,
\end{eqnarray}
where $\mathcal{H}\equiv a'/a$ is the conformal Hubble parameter. The abundance of the GWs evaluated deep inside the radiation-dominated era can be related to the present-day GW abundance as~\cite{2018PhRvD..97j3528A}
\begin{eqnarray}
\Omega_{\rm GW}^{(0)}(k)\,h^2
\simeq 1.62\times10^{-5}\,\Omega_{\rm GW}(k,\eta),.
\end{eqnarray}
For sufficiently large initial PBH abundance ($\beta>\beta_{\rm c}$), the discrete nature of PBHs generates isocurvature density fluctuations. These perturbations, initially Poissonian, are converted into adiabatic modes once PBHs dominate the energy density. Defining the initial isocurvature perturbation as
$S_i(\mathbf{x})\equiv\delta\rho_{\rm BH}(\mathbf{x})/\rho_{\rm BH}$,
its real-space two-point function is given by~\cite{2021JCAP...03..053P}
\begin{eqnarray}
\left\langle S_i(\mathbf{x})S_i(\tilde{\mathbf{x}})\right\rangle
=\frac{4\pi}{3}\left(\frac{d}{a}\right)^3
\delta(\mathbf{x}-\tilde{\mathbf{x}})\,,
\end{eqnarray}
where $d$ denotes the mean PBH separation. In Fourier space, this corresponds to a power spectrum
\begin{eqnarray}
\mathcal{P}_{\rm BH,i}(k)=\frac{2}{3\pi}
\left(\frac{k}{k_{\rm UV}}\right)^3\,,
\end{eqnarray}
with $k_{\rm UV}$ being the ultraviolet cutoff associated with PBH evaporation.

These isocurvature perturbations source the Newtonian potential according to
\begin{eqnarray}
\Phi''+3\mathcal{H}(1+c_s^2)\Phi'
+\left[(1+3c_s^2)\mathcal{H}^2
+2\mathcal{H}'+c_s^2k^2\right]\Phi
=\frac{1}{2}a^2 c_s^2\rho_{\rm BH} S_i\,,
\end{eqnarray}
which induces second-order GWs. Solving the momentum integrals, the GW spectrum near the resonant peak takes the form
\begin{eqnarray}
\Omega_{\rm GW,res}(k)\simeq
\Omega_{\rm GW,res}^{\rm peak}
\left(\frac{k}{k_{\rm UV}}\right)^{11/3}
\Theta^{\rm (iso)}_{\rm UV}(k)\,,
\end{eqnarray}
where the amplitude near the peak can be expressed as 
\begin{eqnarray}
\label{eq:OGWresPeak}
   \Omega_{\rm GW, res}^{\rm peak} = C^4(w)\frac{c_s^{7/3}(c_s^2-1)^2}{576 \times 6^{1/3}\pi}
    \left(\frac{k_{\rm BH}}{k_{\rm UV}}\right)^8
    \left(\frac{k_{\rm UV}}{k_{\rm ev}}\right)^{17/3},
\end{eqnarray}
with $C(w) = \frac{9}{20} \alpha_{\rm fit}^{-\frac{1}{3w}}\left(3+\frac{1-3w}{1+3w}\right)^{-\frac{1}{3w}}$, considering $\alpha_{\rm fit}\approx 0.135$. Here, $w$ presents the background equation-of-state. $\Theta^{\rm (iso)}_{\rm UV}(k)$ is the UV cutoff function for the isocurvature case~\cite{Paul:2025kdd}. The PBH domination scale ($k_{\rm BH}$) is defined as following 
\begin{eqnarray}
\label{eq:kBH}
    k_{\rm BH} &=& \sqrt{2}\gamma^{1/3}\beta^{\frac{1+w}{6w}}k_{\rm UV}.
\end{eqnarray}
For $k\ll k_{\rm UV}$, the GW spectrum lies in the infrared (IR) region which can be expressed as
\begin{eqnarray}
\label{eq:OGWIR}
    \Omega_{\rm GW, IR}(k) 
    &\simeq& 9.03 \times 10^{24}\, C^4(w) \,\beta^{\frac{4(1+w)} {3w}}
    \left(\frac{\gamma}{0.2}\right)^{\frac{8}{3}}
    \left(\frac{M_{\rm in}}{10^4\text{g}}\right)^{\frac{28}{9}}\left(\frac{k}{k_{\rm UV}}\right).
\end{eqnarray}
A transition frequency $ f_{\rm T}$ marks the boundary between resonance peak and IR tail~\cite{Domenech:2020ssp}, which is determined by  
$\Omega_{\rm GW,IR} (f_{\rm T}) = \Omega_{\rm GW,res} (f_{\rm T})$.
Thus, the resultant behaviour of the isocurvature-induced GW spectrum can be expressed analytically as  
\begin{eqnarray}
\label{eq:GW_iso}
    \Omega_{\rm GW, iso}(f) =
    \begin{cases}
        \Omega_{\rm GW,res}(f), & f \geq f_{\rm T}, \\
        \Omega_{\rm GW,IR}(f), & f < f_{\rm T}.
    \end{cases}
\end{eqnarray}

Adiabatic scalar perturbations also generate sizeable induced GWs during PBH reheating. During PBH domination, the gravitational potential remains constant on subhorizon scales. However, the abrupt transition to radiation domination leads to efficient GW production, even for an almost scale-invariant primordial spectrum. We follow Ref.~\cite{Bhaumik:2024qzd,Domenech:2024wao} for this contribution. Assuming adiabatic initial conditions, the primordial curvature power spectrum is parametrized as
\begin{eqnarray}
\mathcal{P}_{\Phi_0}(k)=A_s
\left(\frac{k}{k_*}\right)^{n_s-1}
\Theta(k_{\rm UV}-k)\Theta(k-k_{\rm IR})\,,
\end{eqnarray}
where the UV and IR cutoffs ensure perturbativity and consistency with CMB observations. For modes entering the horizon before PBH domination, the transfer function can be approximated as
\begin{eqnarray}
\Phi_{\rm ad}(k\gg k_{\rm BH})=
A_\Phi(w)
\left(\frac{k}{k_{\rm BH}}\right)^{n(w)}\,,
\end{eqnarray}
with the exponent $n(w)$ and amplitude $A_\Phi(w)$ determined by the background equation of state.

The effective spectral index governing the GW source term is
\begin{eqnarray}
n_{\rm eff}(w)=-\frac{5}{3}+n_s+2n(w)\,,
\end{eqnarray}
which controls the high-frequency behaviour of the induced spectrum. For $k\gg k_{\rm BH}$, the resonant contribution to the GW spectrum reads
\begin{eqnarray}
\Omega_{\rm GW,res}^{\rm ad}(k)=
\Omega_{\rm GW,res}^{\rm ad,peak}
\left(\frac{k}{k_{\rm UV}}\right)^{2n_{\rm eff}(w)+7}
\Theta^{\rm (ad)}_{\rm UV}(k)\,.
\end{eqnarray}
the amplitude is given by
\begin{eqnarray}
    \label{eq:OmegaGWres_Peak_adi}
   \Omega_{\rm GW, res}^{\rm ad, peak} \simeq 9.72\times 10^{32} \frac{ 3^{2 n(w)+n_s}}{4^{3 n(w)+n_s}} A_s^2 A_\Phi(w)^4 \gamma ^{-4 n(w)/3}  \beta ^{-\frac{2 n(w) (w +1)}{3 w }}\left(\frac{g_*(T_{\rm BH})}{108}\right)^{-\frac{17}{9}} \left(\frac{M_{\rm in}}{10^4\text{g}}\right)^{\frac{34}{9}}\,.\nonumber\\
\end{eqnarray}
Here, $\Theta^{\rm (ad)}_{\rm UV} (k)$ is the UV cutoff of the adiabatic case~\cite{Paul:2025kdd}. At the proximity of $k_{\rm BH}$ and at $k\ll k_{\rm BH}$, we can define the intermediate and IR part of the GW spectrum as follows:
\begin{align}    
    \Omega_{\rm GW, mid}^{\rm ad}(k)&=& -\frac{ A_s^2 A_\Phi(w)^4 c_s^4}{768 (1+2n_{\rm eff})}\left(\frac{3}{2}\right)^{\frac{2}{3}} \xi_1^{1+2n_{\rm eff}}
    \left(\frac{k_{\rm BH}}{k_{\rm UV}}\right)^{2 n_s-\frac{7}{3}}
    \left(\frac{k_{\rm UV}}{k_{\rm ev}}\right)^{\frac{14}{3}}
    \left(\frac{k}{k_{\rm UV}}\right)^{5},\label{eq:OmegaGW_mid_adi}\\
    \Omega_{\rm GW, IR}^{\rm ad}(k)&=&\frac{ A_s^2 c_s^4 (3 w +5)^4 }{10^4  (6 n_s+5) (w +1)^4}
    \left(\frac{3}{2}\right)^{\frac{2}{3}}
    \left(\frac{\xi_2 k_{\rm BH}}{k_{\rm UV}}\right)^{\frac{5}{3} + 2n_s}
    \left(\frac{k_{\rm UV}}{k_{\rm ev}}\right)^{\frac{14}{3}}
    \left(\frac{k}{k_{\rm UV}}\right) \label{eq:OmegaGW_IR_adi}.
\end{align}
Here, both $\xi_1$ and $\xi_2$ are the functions of $w$, whose analytical estimations are discussed in Ref.~\cite{Domenech:2024wao,Paul:2025kdd}.
Now, the complete spectrum for the adiabatic contribution of the GW spectra ($\Omega_{\rm GW, ad}(k)$), is  obtained by combining the resonant, intermediate, and IR contributions. We again stress the fact that in the study we consider the production of PBH occurs during radiation-domination epoch. Thus, we set $w=1/3$, in the analysis.

\begin{figure*}
    \centering
    \includegraphics[scale=0.5]{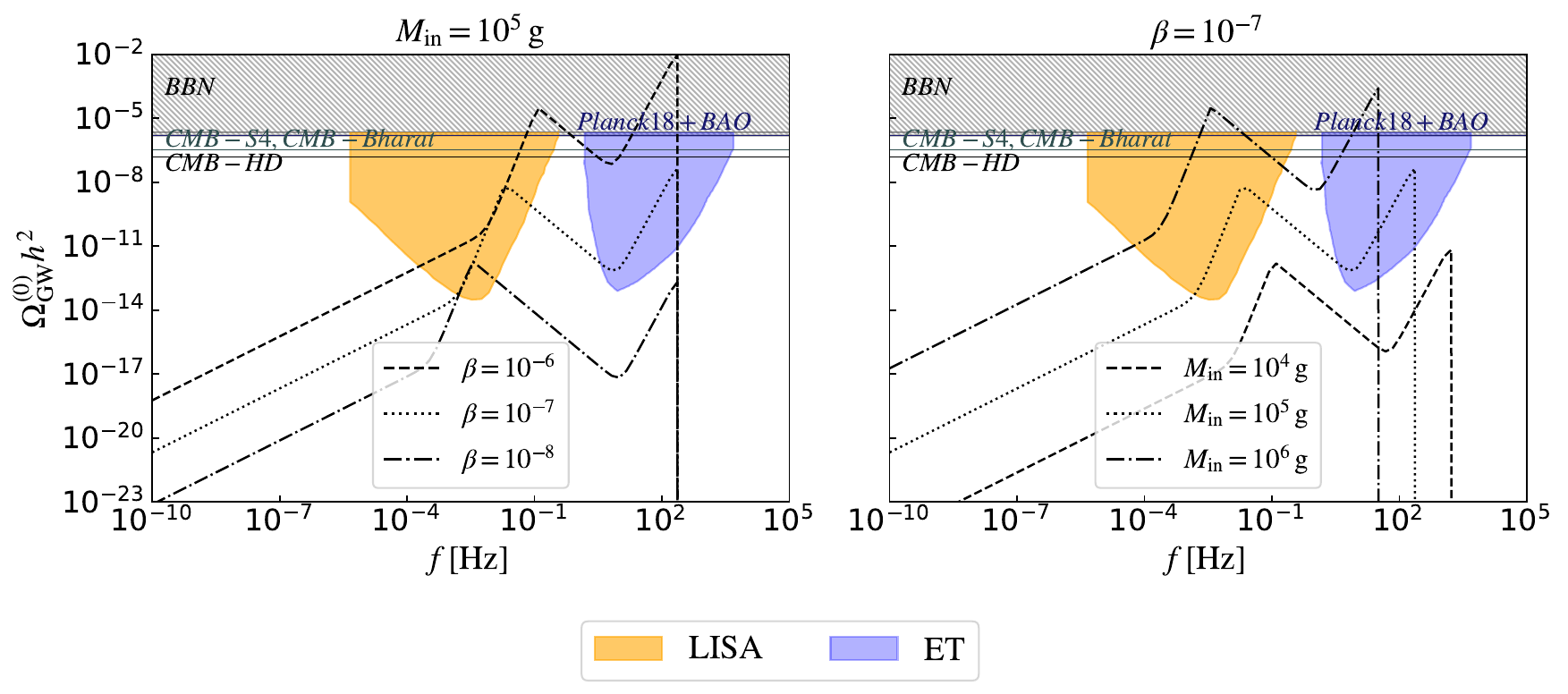}
    \caption{\it Illustration of present-day induced GW spectrum considering adiabatic and isocurvature fluctuation. \textbf{Left panel} shows the variation of GW spectrum with respect to $\beta$, keeping $M_{\rm in}=10^5$ g, while the \textbf{right panel} represents the variation with respect to $M_{\rm in}$ for $\beta=10^{-7}$. Present constraints and projected upper limits on the GW spectra are shown in the panels. Two colour shades indicate the projected sensitivities of LISA and ET.}
    \label{fig:GW_full}
\end{figure*}

The two sources of induced GWs are independent and contribute to the spectrum of the induced GWs simultaneously in the PBH reheating scenario. Thus, under linear perturbation theory, the resulting present-day energy density spectrum of the induced GWs can be expressed as  

\begin{eqnarray}  
\label{eq:GW_combined}  
\Omega_{\rm GW,com}^{(0)}(f) h^2 = \Omega_{\rm GW,iso}^{(0)} (f) h^2 + \Omega_{\rm GW,ad}^{(0)} (f) h^2.  
\end{eqnarray}  
Evaluating the detectability of $\Omega_{\rm GW,com}^{(0)}(f)$, at the future GW detectors, provides a comprehensive understanding of the observational prospects for early Universe physics, and reveal the nature of DM. The resulting combined GW spectrum is displayed in Fig.~\ref{fig:GW_full}. The contribution sourced by adiabatic fluctuations exhibits a peak at comparatively high frequencies, corresponding to the ultraviolet cutoff scale $k_{\rm UV}$, whereas the isocurvature-induced component peaks at the characteristic PBH domination scale $k_{\rm BH}$, which lies at lower frequencies. As a result, the total GW signal features two well-separated peaks at distinct characteristic frequencies, as clearly illustrated in the figure.
This complementary spectral structure significantly broadens the frequency window over which the GW signal is enhanced, thereby improving the prospects for detection by forthcoming GW interferometers. Lighter PBHs form and evaporate at earlier times, which shifts both peaks toward higher frequencies, a trend that is evident in Fig.~\ref{fig:GW_full}.
Furthermore, while the peak frequency is set by the characteristic PBH scale, the amplitude exhibits a distinct dependence on the initial PBH abundance for the two production mechanisms. For a radiation-dominated background ($w=1/3$), the amplitude of the isocurvature-induced GW spectrum scales as $\beta^{16/3}$ (see Eq.~\eqref{eq:OGWresPeak}), whereas the amplitude of the adiabatic-induced GW spectrum follows a steeper scaling, $\beta^{4.33}$ (cf. Eq.~\eqref{eq:OmegaGWres_Peak_adi}).

\begin{figure}[ht!]
    \centering
    \includegraphics[scale=0.35]{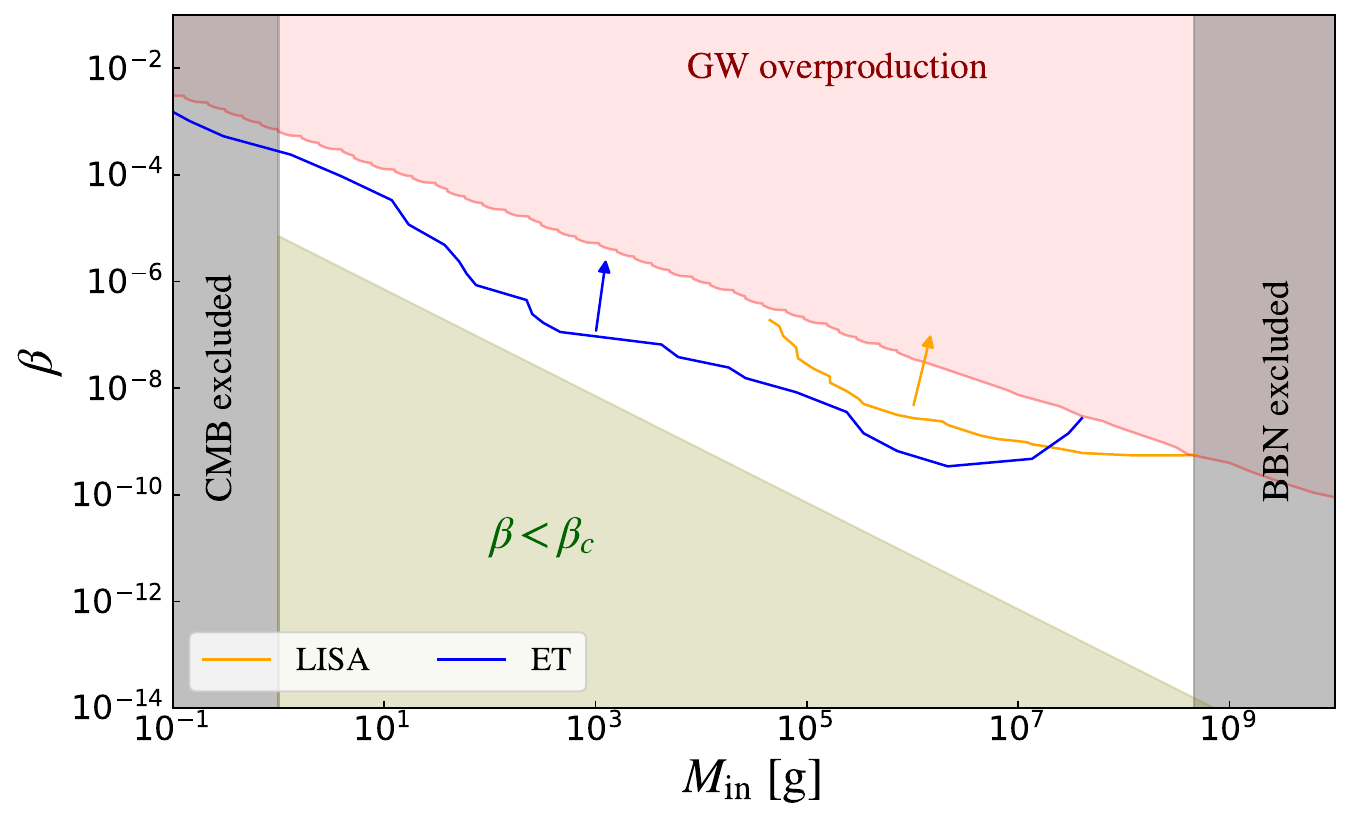}
    \caption{\it Illustration of SNR in $M_{\rm in}-\beta$ plane for LISA and ET. Each curve presents $\rm SNR=1$ contour for the corresponding detector, while the arrows indicate regions where $\rm SNR>1$. The red-hatched region denotes exclusion due to GW overproduction, and the grey-hatched areas are ruled out by CMB and BBN constraints. $\beta<\beta_c$ is highlighted by olive-hatch.}
    \label{fig:snr}
\end{figure}

To asses the detection prospect of any signal, the estimation of signal-to-noise ratio (SNR) is essential. For a GW signal, SNR can be calculated as~\cite{Thrane:2013oya,Caprini:2015zlo}
\begin{eqnarray}
\label{eq:snr}
    \text{SNR}\;\equiv\; \sqrt{T_{\rm obs}\;\int_{f_{\rm min}}^{f_{\rm max}} df \left(\frac{\Omega_{\rm GW}^{(0)}(f) h^2}{\Omega_{\rm GW}^{\rm noise}(f)h^2}\right)^2}\,.
\end{eqnarray}
The quantity $\Omega_{\rm GW}^{\rm noise}(f)\,h^2$ characterizes the detector noise spectrum over the accessible frequency band $[f_{\rm min},f_{\rm max}]$, while $T_{\rm obs}$ represents the total observation time. Table-\ref{tab:detector_spec} presents the frequency band and observation time for LISA~\cite{amaroseoane2017laser,Baker:2019nia} and ET~\cite{Punturo_2010,Hild:2010id}.
\begin{table}[!ht]
    \centering
    \renewcommand{\arraystretch}{1.2}
    \begin{tabular}{|c|c|c|c|}
    \hline
    \hline
       Detectors  & Frequency band [Hz] & $T_{\rm obs}$ [years]\\
    \hline
       LISA  & $\left[10^{-4}-1\right]$ & $4$ \\
       ET  & $\left[1-10^4\right]$ & $5$ \\
    \hline
    \hline
    \end{tabular}
    \caption{\it Total observation time and operating frequency range for LISA and ET.}
    \label{tab:detector_spec}
\end{table}
In our previous work~\cite{Paul:2025kdd}, we demonstrated the detection prospects of the induced GW signal at LISA and ET. As a summary, Fig.~\ref{fig:snr} shows the SNR in the $M_{\rm in}$–$\beta$ plane for both detectors, where each curve represents the $\mathrm{SNR}=1$ contour. For LISA, detection prospects are available in the range $M_{\rm in}\in (10^5,10^7),\mathrm{g}$, whereas for ET the accessible region is significantly broader, covering $M_{\rm in}\in (1,10^7),\mathrm{g}$. Building upon these results, the present study exploits the sensitivity of induced GWs to probe the properties of DM.

\section{Complementary probe of DM with gravitational wave missions}
\label{sec:complementarity}
\begin{figure}[ht!]
    \centering
    \includegraphics[scale=0.5]{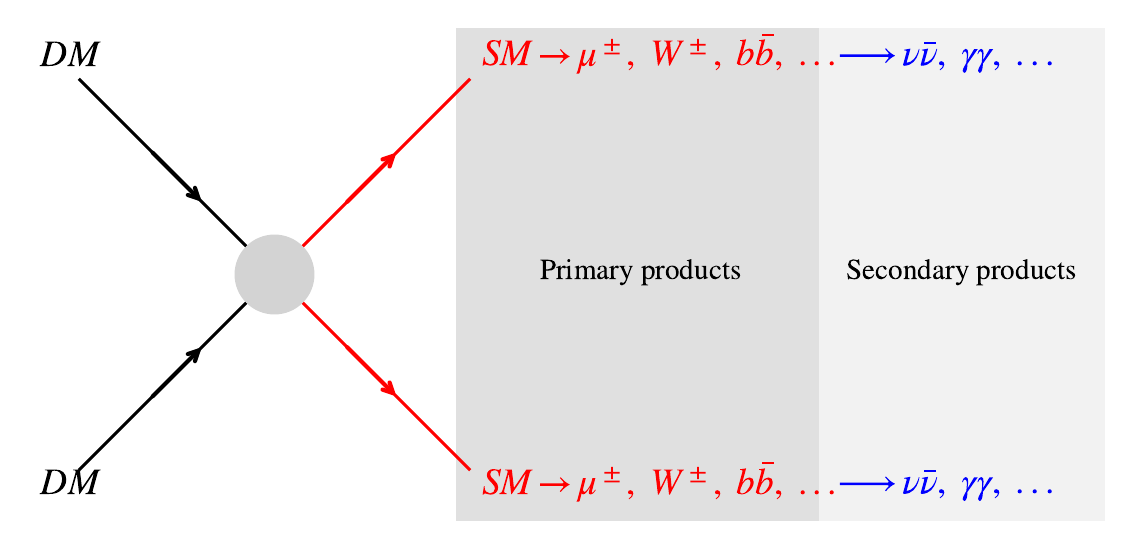}
    \caption{\it Schematic diagram illustrating the annihilation of DM into primary and final particles.}
    \label{fig:feynman_indirect}
\end{figure}

Annihilating DM particles can produce a variety of SM final states, such as quarks, leptons, or electroweak gauge bosons, which subsequently hadronize into secondary particles, most notably neutrinos ($\nu$) and  photon ($\gamma$ rays). In addition, DM may annihilate directly into photons or neutrinos. Such final states constitute the primary observables in indirect detection searches. A schematic illustration of these processes is shown in Fig.~\ref{fig:feynman_indirect}.

The resulting $\gamma$-ray and neutrino spectra can be probed in astrophysical environments with large DM densities, such as the Galactic centre, dwarf spheroidal galaxies, and galaxy clusters. Current $\gamma$-ray experiments, including HESS~\cite{HESS:2021zzm,HESS:2020zwn,HESS:2016mib,HESS:2015cda,HESS:2014zqa}, MAGIC~\cite{MAGIC:2021mog,MAGIC:2020ceg,MAGIC:2016xys,Aleksic:2013xea}, HAWC~\cite{HAWC:2018eaa,HAWC:2017mfa}, FERMI-LAT~\cite{Thorpe-Morgan:2020czg,Hoof:2018hyn,Fermi-LAT:2016uux,Fermi-LAT:2015xij,Fermi-LAT:2015att}, put a level of constraints on DM annihilation into channels such as $b\bar{b}$, $q\bar{q}$, $\mu^+\mu^-$, $W^+W^-$, and $\gamma\gamma$. These bounds are expected to be tighter by upcoming facilities, such as the Cherenkov Telescope Array (CTA)~\cite{CherenkovTelescopeArray:2023aqu}. Further constraints arise from neutrino-based indirect searches, including ANTARES~\cite{ANTARES:2022aoa, Albert:2016emp}, IceCube~\cite{2023PhRvD.108j2004A,2017EPJC...77..627A} and Super-Kamiokande~\cite{Super-Kamiokande:2020sgt,Frankiewicz:2015zma}, which are primarily providing constraints, assuming NFW profile of the galactic halos. Future experiments such as the Cubic Kilometre Neutrino Telescope (KM3NeT)~\cite{KM3NeT:2024xca} are expected to extend the sensitivity to a broader DM mass range. The inferred limits are expected to depend  on both the annihilation channel and the astrophysical modelling of DM distribution inside the halos. 
GW experiments, in contrast, are largely unaffected by uncertainties associated with the modelling of late-time astrophysical environments~\footnote{The detection of a stochastic GW background does require the subtraction of astrophysical foregrounds, which remains an active area of research.}. Since primordial gravitational waves  propagate essentially unimpeded after their production, GW observations can help in having a cleaner probe of the thermal and expansion history of the early Universe. In the presence of a PBH-dominated epoch, the altered thermal history modifies the DM parameter space consistent with the observed relic abundance. Because the GWs spectrum is likewise sensitive to the same expansion history, GW observations provide an indirect and complementary probe of DM properties such as its mass and annihilation cross-section.

By \emph{complementary probes of DM}, we specifically mean the ability of two otherwise distinct observational avenues, indirect detection of DM and GW observations, to access the same region of DM parameter space, characterized by the mass and annihilation cross-section of DM. Aforementioned sections establishes the relations the PBH-parameters ($\beta$, $M_{\rm in}$), responsible for early matter domination, are related with DM parameters ($\mx$, $\langle\sigma v\rangle$). As the GW experiments are strongly depend on the PBH-parameters, the relation allows us to test DM scenario through GW missions, along with traditional indirect searches. To quantify this complementarity, we estimate the SNR analysis for the future GW missions to illustrate how the upcoming missions can probe DM parameters. We focus on two representative production mechanisms, thermal freeze-in and freeze-out, highlighting the regions of parameter space that can be jointly or uniquely tested by GW missions.

\subsection{Freeze-in DM: unique sensitivity of GW searches}

We first consider the freeze-in production of DM in the presence of PBH domination. Since DM-SM interaction is extremely feeble in this case, the annihilation cross-section, $\langle\sigma v\rangle$, is far below the reach of conventional indirect detection experiments. As a result, alternative probes are essential.

\begin{figure*}[ht!]
    \centering
    \includegraphics[scale=0.32]{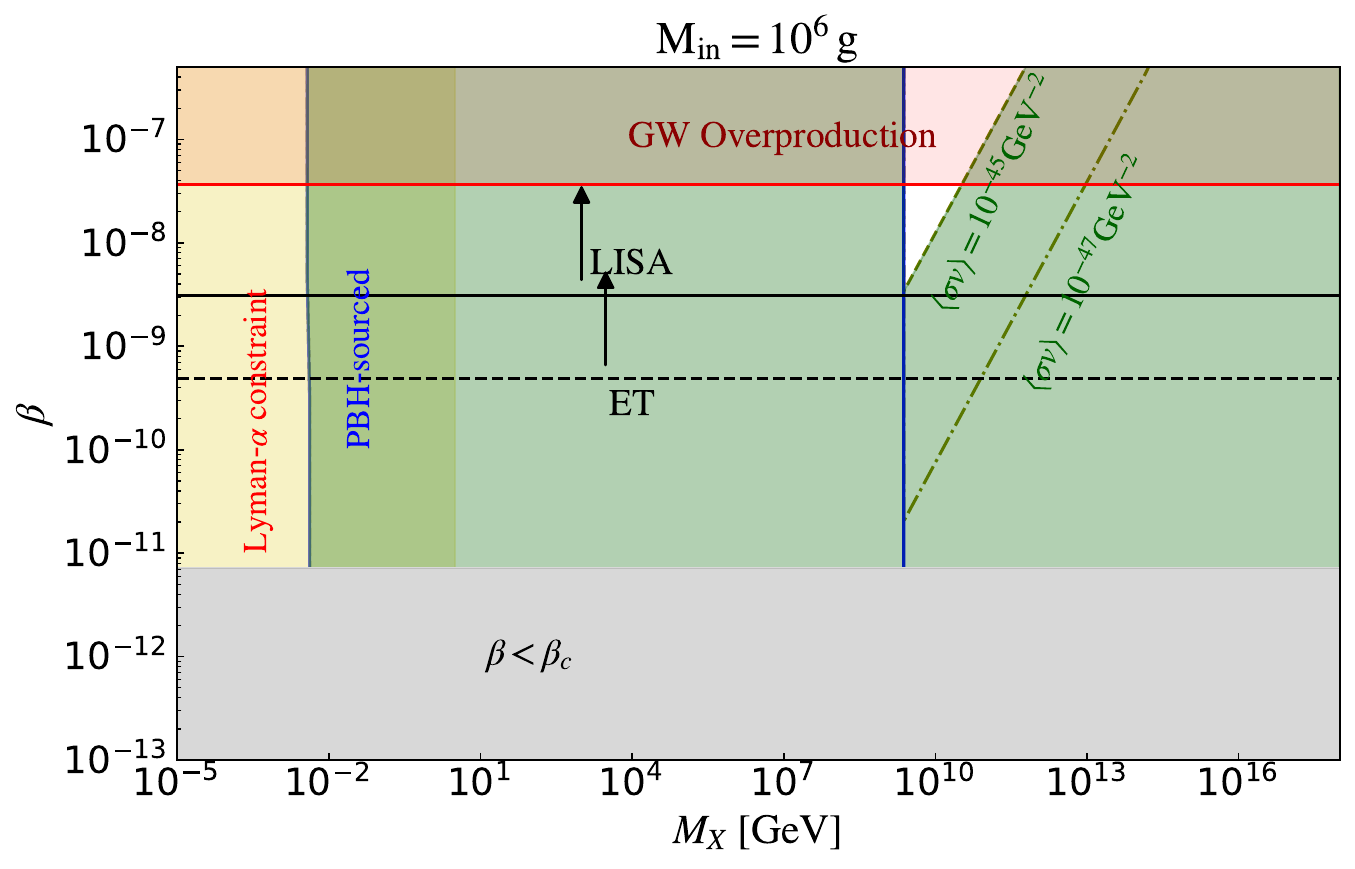}
    \includegraphics[scale=0.32]{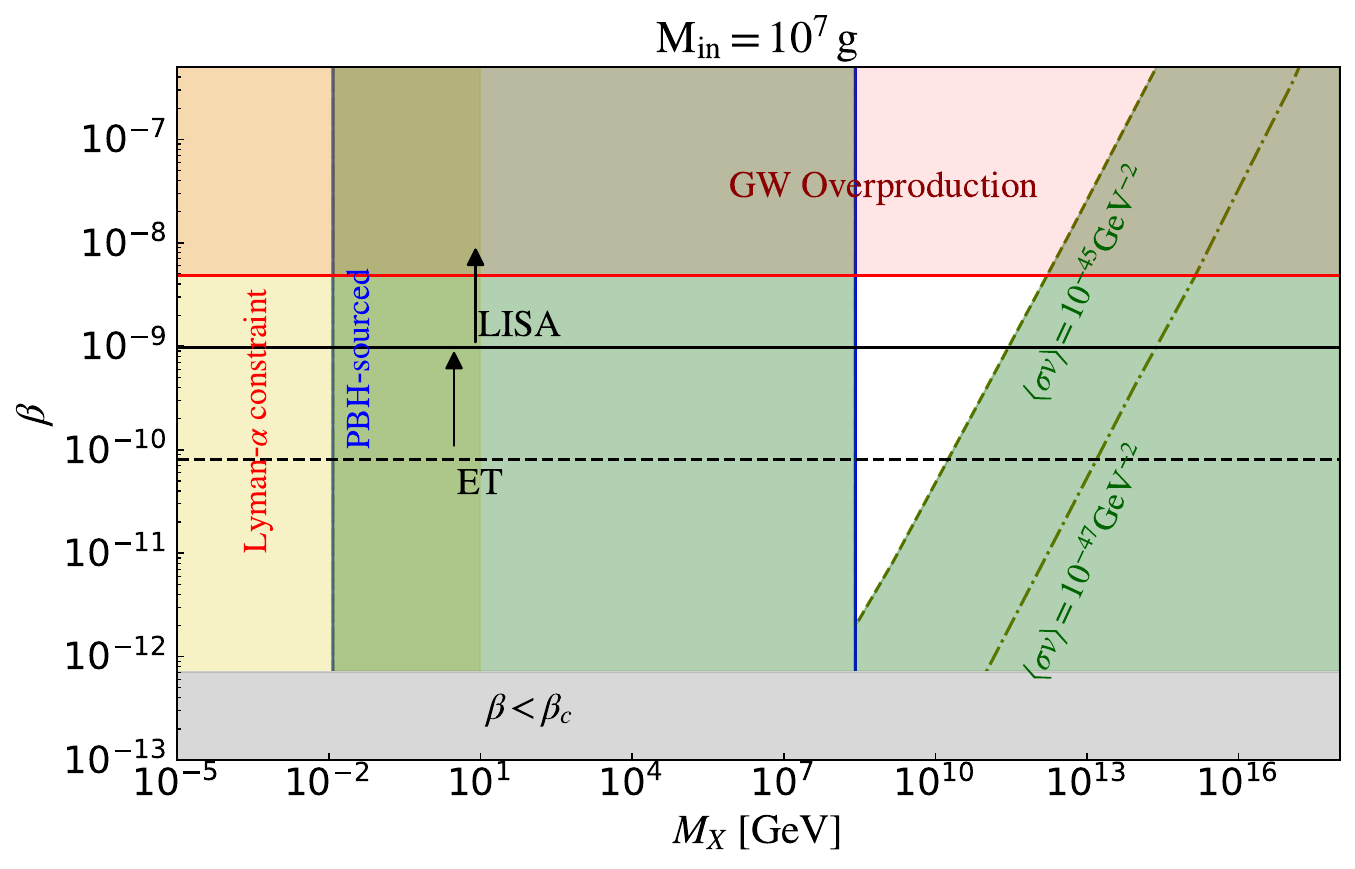}
    \caption{\it Allowed parameter in $M_X$-$\beta$ plane, incorporating the production of DM from PBH in addition to freeze-in production of DM. Olive-shaded region is excluded due to the overproduction of DM, while white region is allowed. A larger interaction cross-section reduces the allowed region even more. Light-orange and grey-hatched regions are excluded due to GW-overproduction and $\beta<\beta_c$, respectively. SNR $=1$ contour for LISA (black solid) and ET (black dotted) are also presented, with the arrows indicating the SNR $>1$ regions. Lower massive DM are excluded by Lyman-$\alpha$ constraint. \textbf{Left panel} present $M_{\rm in}=10^6$ g, while the \textbf{right} one is for $M_{\rm in}=10^7$ g.}
    \label{fig:MX_beta}
\end{figure*}

\begin{figure}[ht!]
    \centering
    \includegraphics[scale=0.35]{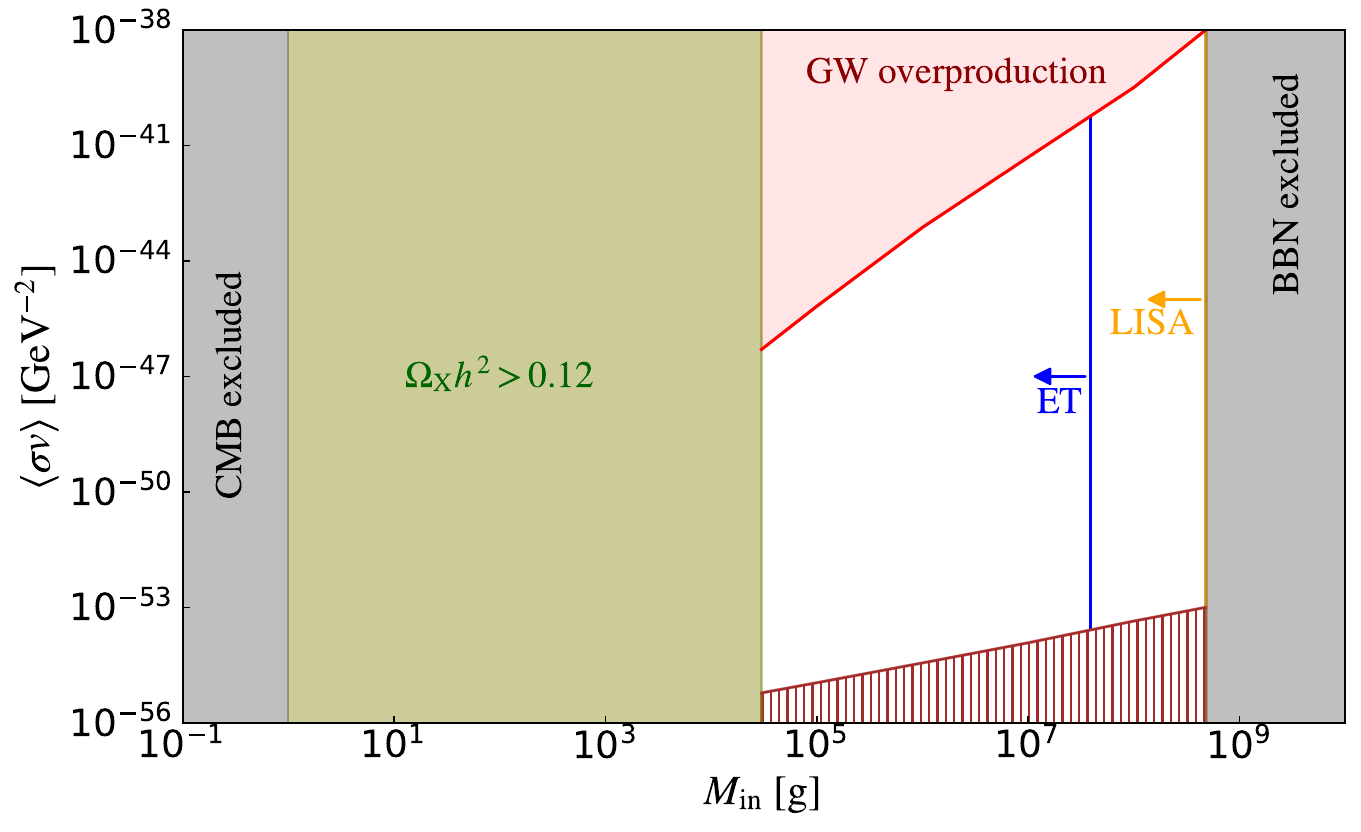}
    \caption{\it Allowed range of $\langle \sigma v\rangle$, considering PBH-sourced DM, including the contribution from thermal freeze-in production, in the $M_{\rm in}$–$\langle\sigma v\rangle$ plane. The red-hatched region is excluded due to the overproduction of GWs. The green-shaded region corresponds to the overproduction of DM. The brown-hatched region denotes the parameter space where the required DM mass exceeds the Planck scale, $\mx > M_P$, in order to satisfy the observed relic abundance. The vertical blue and orange lines, along with their respective arrows, mark the regions accessible to ET and LISA, respectively. The regions excluded by BBN and CMB constraints are indicated by grey-hatched areas.
}
    \label{fig:sigmav_max}
\end{figure}

Fig.~\ref{fig:MX_beta} shows the viable parameter space in the $\mx-\beta$ plane after incorporating both PBH-sourced and freeze-in production of DM, together with the constraints from induced GWs associated with PBH reheating, including bounds from GW overproduction as constrained by $\Delta N_{\rm eff}$ bound at BBN. The observed relic abundance can be achieved over a wide range of DM masses and annihilation cross-sections for PBH parameters $(\beta, M_{\rm in})$. The viable regions are determined by the interplay between entropy dilution during the PBH-dominated epoch, controlled by $(\beta, M_{\rm in})$ and the efficiency of freeze-in production governed by $\langle \sigma v \rangle$.

As already mentioned, the entire range of $\langle\sigma v\rangle$ compatible with freeze-in lies well below the sensitivity of existing and near-future indirect detection experiments. However, these same regions can be constrained through GW observations. In particular, the requirement that PBH-induced gravitational waves do not overproduce the stochastic GW background imposes an upper bound on $\langle\sigma v\rangle$ for each value of $M_{\rm in}$, as shown in Fig.~\ref{fig:sigmav_max}. In addition, the Planck scale $M_P$, which sets the maximum allowed DM mass, establishes a fundamental lower bound on $\langle\sigma v\rangle$. Furthermore, future GW observatories such as LISA and ET  indirectly will be able to probe regions of parameter space corresponding to a signal-to-noise ratio $\rm SNR>1$. The detectable ranges correspond to $\langle \sigma v\rangle \in (10^{-55},\,10^{-40})\,{\rm GeV^{-2}}$ for ET and $\langle \sigma v\rangle \in (10^{-55},\,10^{-38})\,{\rm GeV^{-2}}$ for LISA, respectively.

These results demonstrate that GW searches provide a indirect hints on DM properties, including its mass and interaction strength, and constitute an essentially irreducible probe of freeze-in DM scenarios in PBH-dominated cosmologies. Across large regions of parameter space, gravitational waves offer  one of the viable observational window into the DM interaction strength.

\subsection{Freeze-out DM: complementarity with indirect searches}

We now turn to the freeze-out production of DM, where the interaction strength with the SM bath is sufficiently large for DM to reach thermal equilibrium. In this case, indirect detection experiments regain sensitivity, allowing for a complementarity between particle-based searches and GW observations.

\begin{figure*}[ht!]
    \centering
    \includegraphics[scale=0.31]{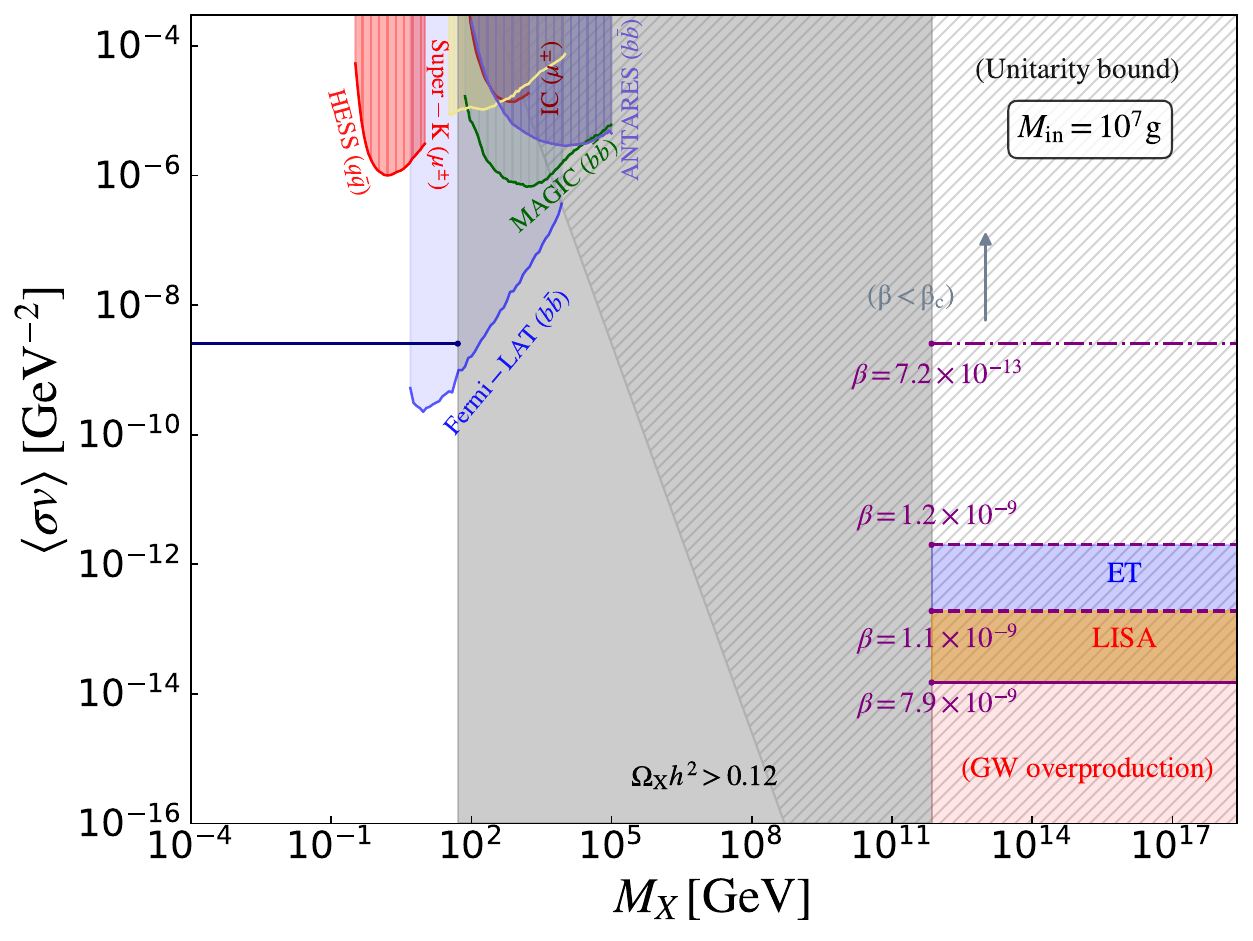}
    \includegraphics[scale=0.31]{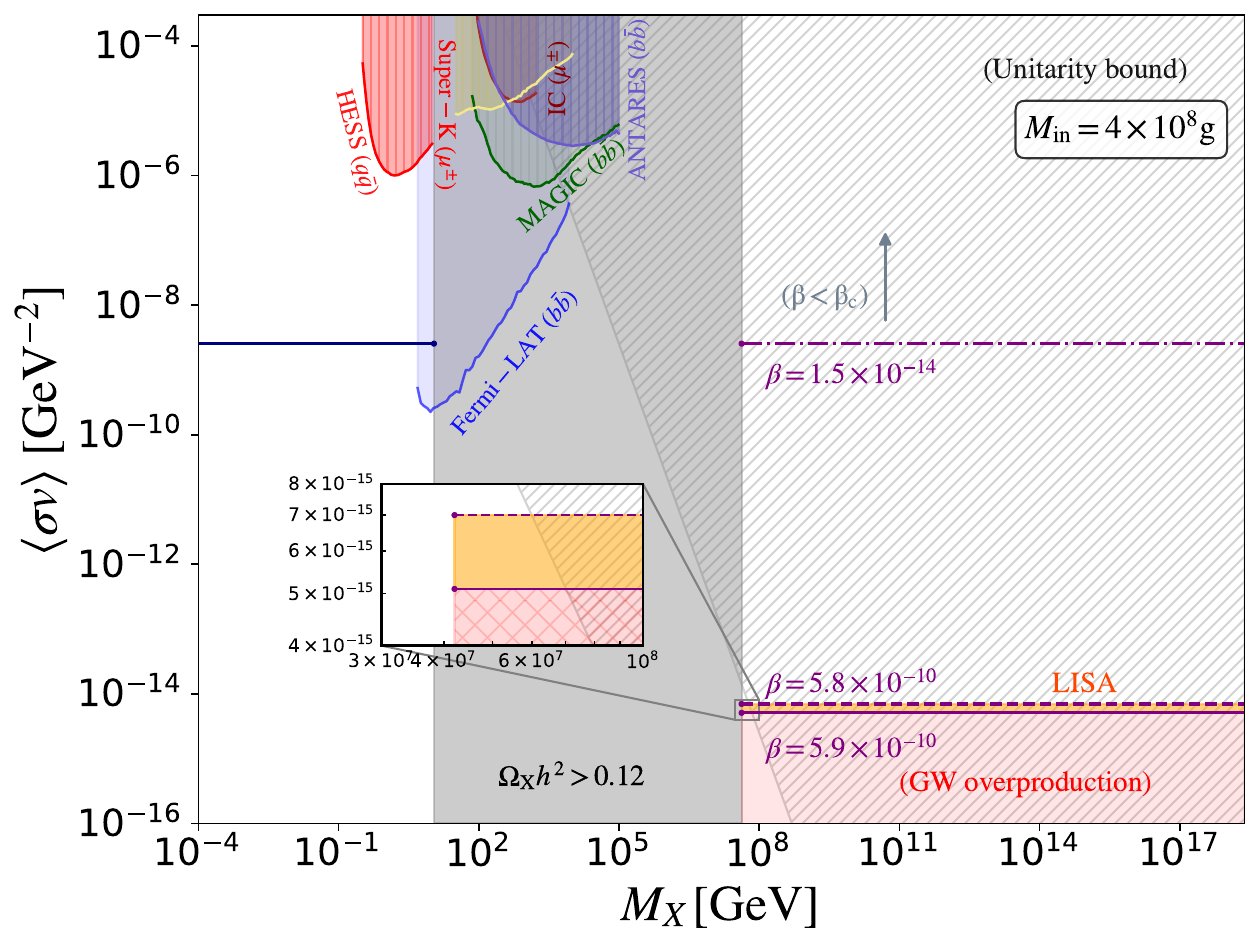}
    \caption{\it Illustration of current experimental constraints from DM indirect searches in $\mx$-$\langle\sigma v\rangle$ plane, for different $M_{\rm in}$. The vertical bands present current constraints from DM indirect searches via $\gamma$-ray and neutrino detection. Different purple horizontal lines presents different values of $\beta$. The horizontal blue and orange bands present $\rm SNR >1$ for LISA and ET, respectively to demonstrate the complementarity between indirect detection and GW experiments. Experimental constraints of the indirect detections are adopted from \href{https://github.com/moritzhuetten/dmbounds/tree/main?tab=readme-ov-file}{\texttt{github/moritzhuetten}}.}
    \label{fig:MX_beta_fo_comp}
\end{figure*}

Fig.~\ref{fig:MX_beta_fo_comp} summarizes the current constraints from DM indirect detection via $\gamma$ rays and neutrinos in the $\mx$--$\langle\sigma v\rangle$ plane for two representative initial PBH masses, $M_{\rm in}=10^7\,{\rm g}$ (left panel) and $M_{\rm in}=4\times10^8\,{\rm g}$ (right panel). We consider two benchmark annihilation channels, $q\bar q$ and $b\bar b$. Constraints from FERMI-LAT, HESS, and MAGIC are shown for $\gamma$-ray searches, while ANTARES, Super-Kamiokande and IceCube bounds are included for neutrino searches. The grey-hatched region in each panel corresponds to the violation of the unitarity bound on the annihilation cross-section, $\langle\sigma v\rangle \gtrsim  8\pi/\mx^2$~\cite{Giudice:2000ex}. The vertical grey band denotes the region of DM masses for which PBH-sourced production alone always leads to an overabundant relic. For DM masses lying to the left of this band, the correct relic abundance is obtained for the standard thermal freeze-out cross-section, $\langle\sigma v\rangle \simeq 2.5\times10^{-9}\,{\rm GeV}^{-2}$. This occurs because, in such scenarios, freeze-out always takes place during radiation domination after the completion of PBH evaporation, i.e., $T_{\rm f}<T_{\rm ev}$. In this case, any DM produced from evaporating PBHs reaches thermal equilibrium due to the large value of $\langle \sigma v \rangle$, and the observed relic abundance is reproduced for the standard WIMP cross-section. Moreover, this result is independent of the $\beta$ value, provided that freeze-out occurs during the standard radiation-dominated epoch after PBH evaporation. Therefore, this region cannot be probed by GW missions, since their sensitivity depends on the value of $\beta$.

In contrast, for DM masses on the right side of the grey band, PBHs are initially underabundant, freeze-out occurs before PBH domination, and the DM yield is subsequently diluted by entropy injection during PBH evaporation. Consequently, larger values of $\beta$, corresponding to a longer PBH-dominated epoch, require smaller annihilation cross-sections to reproduce the observed relic abundance. For each $\beta>\beta_c$, a unique value of $\langle\sigma v\rangle$ is therefore selected, while for $\beta\leq\beta_c$ the required cross-section asymptotes to the standard WIMP value. In both panels, the red-hatched regions indicate GW overproduction, excluded by bounds on extra radiation. The horizontal orange and blue bands highlight the regions detectable by LISA and ET, respectively. However, in most cases, the regions that can be probed by LISA and ET are restricted by the unitarity bound. We find that, for $M_{\rm in}=4\times10^8\,{\rm g}$, only a narrow DM mass window, $\mx\in[4.1\times10^7,\,7\times10^7]\,{\rm GeV}$, with $\langle \sigma v\rangle \in (5\times 10^{-15},\,7\times 10^{-15})\, {\rm GeV^{-2}}$, can be probed by LISA. Therefore, there is no overlapping region that can be probed by both indirect detection searches and GW observations.

\begin{figure*}[ht!]
    \centering
    \includegraphics[scale=0.31]{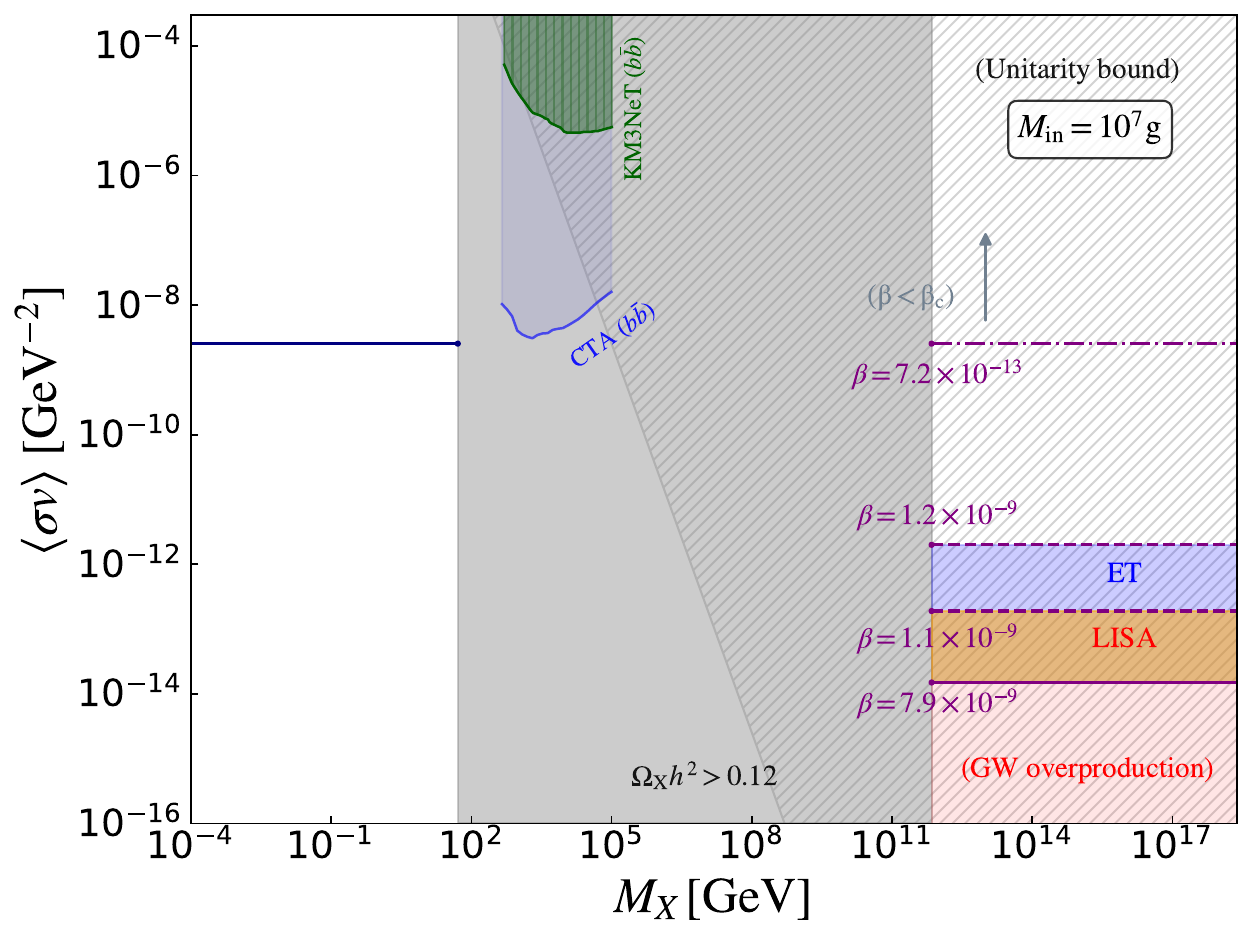}
    \includegraphics[scale=0.31]{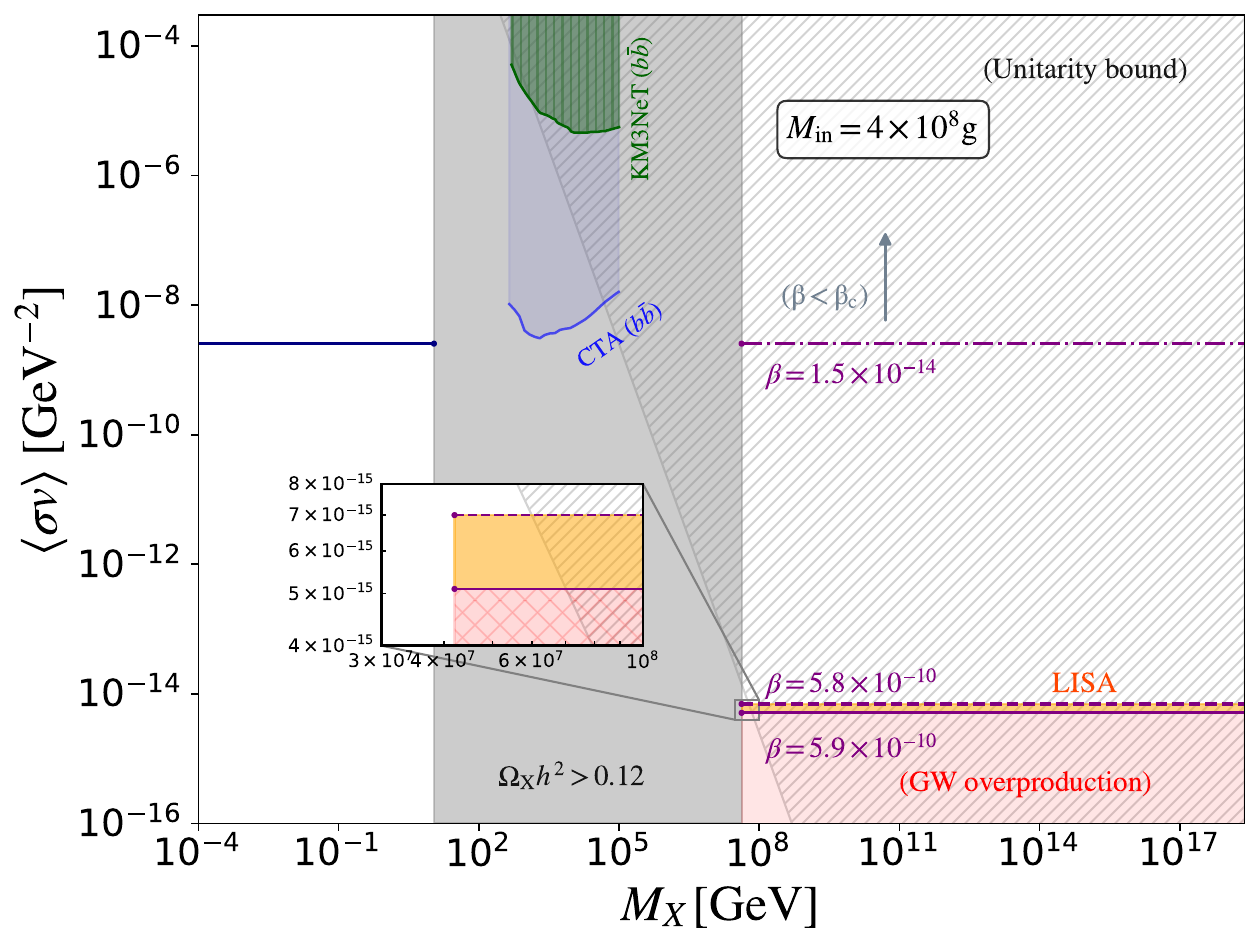}
    \caption{\it Illustration of future projection of $\gamma$-ray and $\nu$-experiments for DM indirect searches. The horizontal blue and orange bands present $\rm SNR >1$ for ET and LISA, respectively, to demonstrate the complementarity between DM indirect searches and GW searches. }
    \label{fig:MX_beta_fo_future}
\end{figure*}

In Fig.~\ref{fig:MX_beta_fo_future}, we extend this analysis to future indirect detection experiments and GW observatories. We include the projected sensitivities of CTA for $\gamma$-ray searches, and KM3NeT for neutrino-based searches. CTA is expected to probe DM masses in the range $\mx\in(200,10^5]\,{\rm GeV}$, while KM3NeT is sensitive to $\mx\in(51,3\times10^4)\,{\rm GeV}$ and $\mx\in(490,10^5]\,{\rm GeV}$, respectively. This figure illustrates the combined future reach of indirect searches and GW observations. Similar to the previous scenario, we find that there is no overlapping region accessible to both indirect detection searches and GW missions. As a result, our goal of demonstrating complementarity between these searches is not realized in this case. In certain DM mass ranges, indirect detection searches remain relevant, whereas at higher DM masses, GW missions emerge as the one of the viable indirect probe.

\begin{figure*}[ht!]
    \centering
    \includegraphics[scale=0.31]{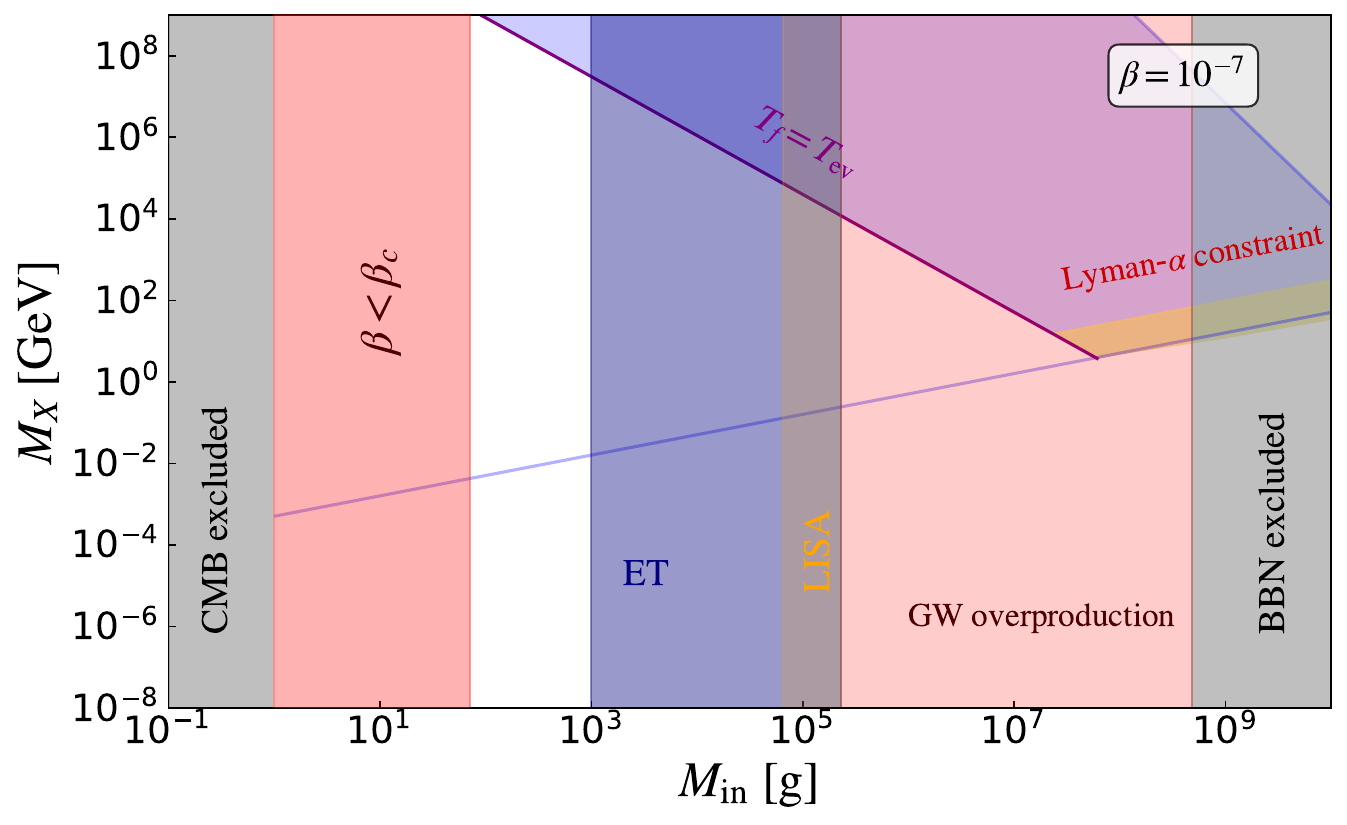}
    \includegraphics[scale=0.31]{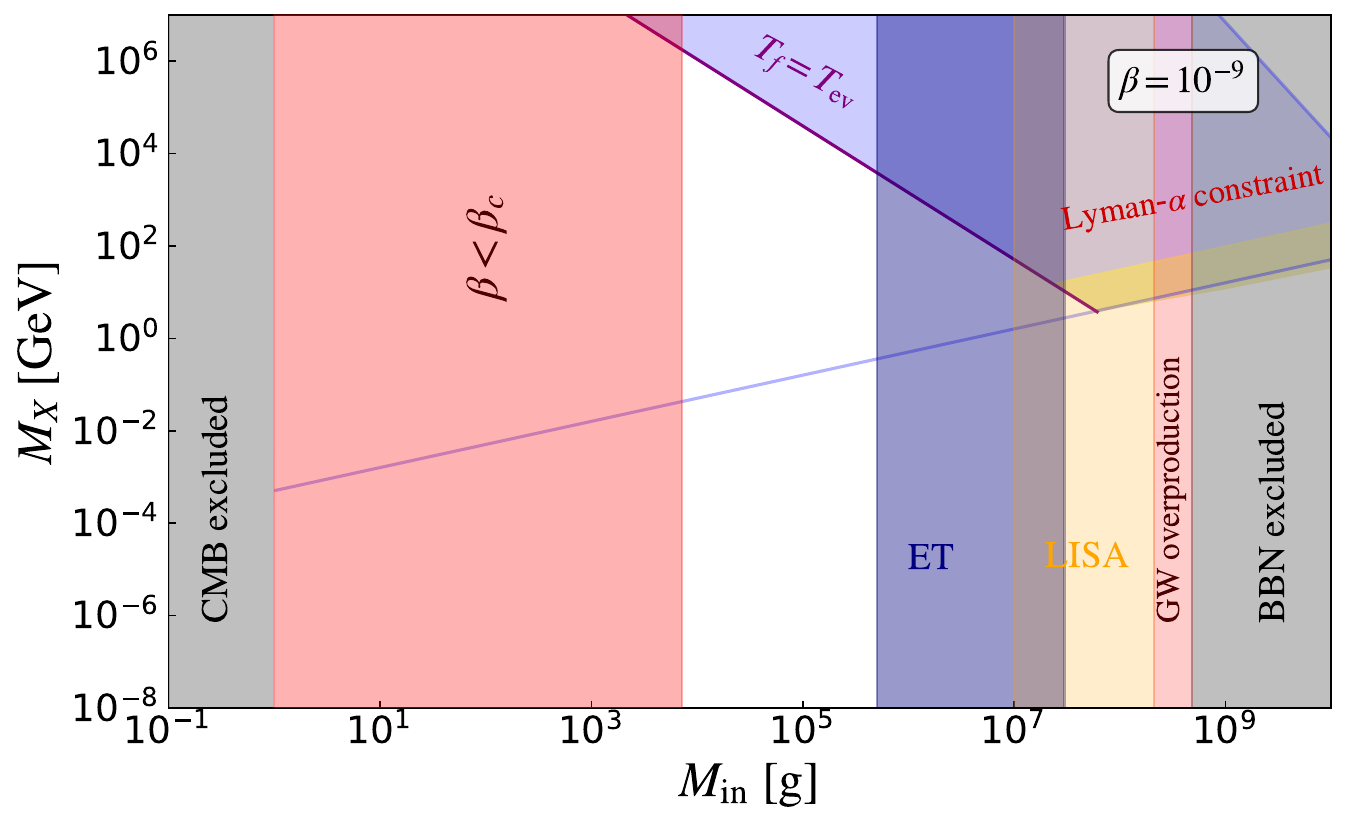}
    \caption{\it Same level of information as in Fig.~\ref{fig:MBH_MX_fo}, with the projections of GW experiments. Blue and orange vertical bands present the projections of ET and LISA, respectively. Light red-hatched region indicates GW-overproduction.}
    \label{fig:MX_beta_fo_gw}
\end{figure*}

In addition, Fig.~\ref{fig:MX_beta_fo_gw} shows the corresponding GW sensitivity in the $M_{\rm in}$--$\mx$ plane for two representative values of $\beta$, with the GW-overproduction regions explicitly indicated. We find that ET can probe PBH masses in the range $M_{\rm in}\in[5\times10^5,\,3\times10^7]\,{\rm g}$, while LISA is sensitive to $M_{\rm in}\in[10^7,\,2.1\times10^8]\,{\rm g}$ for $\beta=10^{-9}$. However, this region is sensitive to the PBH abundance. 

In a summery, these results demonstrate the complementary roles of DM indirect detection searches and GW observations in PBH-dominated cosmologies. While indirect detection constrains the annihilation cross-section through late-time astrophysical signals, GW experiments probe the DM parameters through their imprint on the early Universe thermal history, enabling a genuinely multi-messenger test of DM production mechanisms. However, the same DM parameter space, $(\mx,\langle \sigma v\rangle)$, cannot be simultaneously probed by indirect detection searches using neutrino and $\gamma$-ray observations and future GW missions such as LISA and ET.

\section{Conclusion}
\label{sec:conclusion}

In this work, we present a comprehensive analysis of DM production in a cosmological scenario where the early Universe undergoes a period of PBH domination followed by reheating through Hawking evaporation. Building upon our earlier analysis of induced GWs in PBH reheating \cite{Paul:2025kdd}, we investigate how such a non-standard thermal history impacts DM genesis and its observational signatures.
We systematically analyse several well-motivated DM production channels, including direct production from PBH evaporation, gravitational particle production via graviton-mediated scattering, and thermal production through freeze-in and freeze-out mechanisms. By numerically solving the coupled Boltzmann equations governing the evolution of radiation, PBHs and DM, we identify the regions of parameter space consistent with the observed relic abundance while respecting cosmological constraints from BBN, CMB and Lyman-$\alpha$ forest observations. 

Our results indicate that PBH-evaporation typically dominates production of DM across a broad mass range $M_X \sim 10^{7}\,\mathrm{GeV}$ to $\Mp$, for PBH-formation masses lies $M_{\rm in} \in[ 10^{4} - 4.8\times10^{8}]\,\mathrm{g}$. Conversely, lighter DM candidates are strongly constrained by Lyman-$\alpha$ forest observations (\textit{cf. }Fig.~\ref{fig:MBH_MX_onlyPBH}). While gravitational freeze-in is an inherent process, we find it remains subdominant in the PBH-dominated regime due to substantial entropy injection during the evaporation phase, which dilutes the gravitational contribution (as illustrated in Fig.~\ref{fig:grav_prod}).

Besides the gravitational channel, DM can also be produced through annihilation processes from the thermal bath via both freeze-in and freeze-out mechanisms. The inclusion of thermal production mechanisms, freeze-in and freeze-out substantially modifies the viable parameter space (\textit{c.f.} Fig.~\ref{fig:MBH_MX} and Fig.~\ref{fig:MBH_MX_fo}). If we assume thermal freeze-in production of DM, the total relic abundance receives contributions from three channels, $\Omega_X h^2 = \Omega_X^{\rm PBH} h^2 + \Omega_X^{\rm FI} h^2 + \Omega_X^{\rm grav} h^2 \simeq \Omega_X^{\rm PBH} h^2 + \Omega_X^{\rm FI} h^2$, where the gravitational contribution remains subdominant. We find that thermal freeze-in can compensate for the underabundance of solely PBH-sourced DM for $\mx\gtrsim 10^7$ GeV for heavier PBH, though it simultaneously reduces the parameter space for heavier DM masses due to overproduction risks. On the other hand, in the freeze-out case, the phenomenology depends on whether DM decoupling occurs before or after the completion of PBH evaporation. This behaviour is regulated by the mass of DM and PBH parameters, such as the formation mass and the duration of the PBH-dominated epoch, which is controlled by the parameter $\beta$. Another interesting outcome of this analysis is the freeze-out process is this further allows the parameter regions which were overabundant due to solely PBH-sourced production, for $T_f\geq T_{\rm ev}$ (\textit{cf.} Fig.~\ref{fig:MBH_MX_fo}). Since, the freeze-out occurs after the PBH-evaporation, DM particles thermalize with the plasma and relic abundance is determined by the standard freeze-out mechanism, for $\langle \sigma v\rangle=2.5\times 10^{-9}$ GeV$^{-2}$. 

A central outcome of our analysis is the complementarity between indirect searches of DM and future GW observations. Our analysis focuses on the present and upcoming $\gamma$-ray and neutrino experiments for the indirect searches of DM. On the other hand, PBH-reheating generically produces induced GW backgrounds sourced by both isocurvature fluctuations, arising from the discrete PBH distribution, and adiabatic curvature perturbations, amplified during the transition from PBH domination to radiation-domination. These contributions lead to characteristic doubly peaked GW spectra whose amplitudes and peak frequencies encode information about the PBH mass, initial abundance, and expansion history of the Universe. We show that due to the feeble interaction cross-section for the freeze-in mechanisms, GW missions are the unique probes for them, as $\langle \sigma v\rangle$ lies below the sensitivity of conventional indirect detection experiments. The requirement that PBH-induced gravitational waves do not overproduce the stochastic GW background imposes an upper bound on $\langle\sigma v\rangle$ for each value of $M_{\rm in}$, while the Planck scale $M_P$ sets a fundamental lower bound. Future GW observatories such as LISA and ET can indirectly probe this parameter space through $\rm SNR>1$, with sensitivity to $\langle \sigma v\rangle \in (10^{-55},\,10^{-40})\,{\rm GeV^{-2}}$ for ET and $\langle \sigma v\rangle \in (10^{-55},\,10^{-38})\,{\rm GeV^{-2}}$ for LISA (see, for instance Fig.~\ref{fig:MX_beta}).

We have also investigated the indirect detection prospects of freeze-out DM in PBH-dominated cosmologies using both conventional astrophysical searches and GW observations. Indirect detection experiments probe DM masses up to $\mx \sim 10^5\,{\rm GeV}$, whereas GW observatories extend the sensitivity to much heavier masses  as shown in Figs.~\ref{fig:MX_beta_fo_comp} and \ref{fig:MX_beta_fo_gw}. For instance, for $M_{\rm in}=4\times10^8\,{\rm g}$, LISA can probe a narrow window $\mx\in[4.1\times10^7,\,7\times10^7]\,{\rm GeV}$ with $\langle\sigma v\rangle\sim(5$–$7)\times10^{-15}\,{\rm GeV^{-2}}$. However, a significant fraction of the parameter space is restricted by the unitarity bound, $\langle\sigma v\rangle \lesssim 8\pi/\mx^2$, which further limits the viable region. Consequently, the regions accessible to GW missions and indirect detection experiments do not overlap, establishing GW observations in some cases, unique probe of heavy DM and PBH reheating. This highlights the crucial role of GW observations in probing otherwise inaccessible regions of the DM parameter space beyond the reach of conventional indirect detection experiments.

Overall, our results demonstrate that PBH reheating provides a viable framework for studying the connection between early Universe cosmology, DM production, and GW signals. Indirect detection experiments and next-generation GW observatories probe different regions of the DM parameter space, with indirect searches primarily sensitive to lower DM masses and larger annihilation cross-sections, while GW observations extend the reach to heavier DM masses and smaller interaction strengths. In particular, GW observations provide access to regions that are difficult to probe using conventional indirect detection methods. These findings highlight the potential role of future GW measurements in improving our understanding of DM production and the thermal history of the early Universe.


\begin{acknowledgments}
We acknowledge the computational facilities of Technology Innovation Hub, Indian Statistical Institute (ISI), Kolkata. DP thanks ISI, Kolkata for financial support through Senior Research Fellowship.
SP thanks  ANRF, Govt. of India for partial support through Project No.
CRG/2023/003984.
\end{acknowledgments}

\bibliographystyle{bibi}
\bibliography{mybib.bib}

\end{document}